\documentclass[useAMS,usenatbib,usegraphicx]{mn2e}
\usepackage{graphicx, amssymb, epsfig, fixltx2e, times, color, ulem, upgreek}
\usepackage[fleqn]{amsmath}

\DeclareSymbolFont{UPM}{U}{eur}{m}{n}
\DeclareMathSymbol{\partialup}{0}{UPM}{"40}
\def\kms{\mbox{km~s$^{-1}$}}
\def\kpc{\mbox{kpc}}

\def\Mvir{\mbox{$M_{\rm vir}$}}
\def\Msun{\mbox{$\rm M_{\sun}$}}
\def\rmax{\mbox{$r_{\rm max}$}}
\def\rt{\mbox{$r_{\rm t}$}}
\def\rs{\mbox{$r_{\rm s}$}}
\def\rtrs{\mbox{$r_{\rm t} / r_{\rm s}$}}
\def\vcirc{\mbox{$v_{\rm circ}$}}
\def\vmax{\mbox{$v_{\rm max}$}}
\def\fb{\mbox{$f_{\rm b}$}}
\def\rmi{{\rm i}}
\def\rmf{{\rm f}}
\newcommand{\HI}{\hbox{{\rm H}\kern 0.1em{\sc i}}}

\title[Effects of baryon removal]{Effects of baryon removal on the structure of dwarf spheroidal galaxies}
  \author[K. S. Arraki et al.]{
  Kenza~S.~Arraki$^{1}$ \thanks{NSF Graduate Research Fellow.} \thanks{E-mail: karraki@nmsu.edu}, 
  Anatoly~Klypin$^{1}$, Surhud~More$^{2}$ and Sebastian~Trujillo-Gomez$^{1}$\\
  $^1$ Astronomy Department, New Mexico State University, MSC 4500, PO Box 30001, Las Cruces, NM 88003-8001, USA\\
  $^2$ Kavli Institute for the Physics and Mathematics of the Universe (WPI), University of Tokyo, 5-1-5 Kashiwanoha, Kashiwa, Chiba 277 8583, Japan \\
}
\begin{document}

\date{Accepted 2013 November 25. Received 2013 August 29; in original form 2012 December 22}
\pagerange{\pageref{firstpage}--\pageref{lastpage}} \pubyear{2013}
\maketitle
\label{firstpage}

\begin{abstract}
Dwarf spheroidal galaxies (dSphs) are extremely gas-poor, dark matter-dominated galaxies, which make them ideal to test the predictions of the cold dark matter (CDM) model.  We argue that the removal of the baryonic component from gas-rich dwarf irregular galaxies, the progenitors of dSphs, can substantially reduce their central density. Thus, it may play an important role in alleviating one of the problems of the CDM model related to the structure of relatively massive satellite galaxies of the Milky Way (MW). Traditionally, collisionless cosmological {\it N}-body simulations are used when confronting theoretical predictions with observations. However, these simulations assume that the baryon fraction everywhere in the Universe is equal to the cosmic mean, an assumption which is incorrect for dSphs. We find that the combination of (i) the lower baryon fraction in dSphs compared to the cosmic mean and (ii) the concentration of baryons in the inner part of the MW halo can go a long way towards explaining the observed circular velocity profiles of dSphs. We perform controlled numerical simulations that mimic the effects of baryons. From these we find that the blowing away of baryons by ram pressure, when the dwarfs fall into larger galaxies, decreases the circular velocity profile of the satellite and reduces the density in the central $\sim$200--500~pc by a factor of $(1-\fb)^4\approx 0.5$, where $\fb$ is the cosmological fraction of baryons. Additionally, the enhanced baryonic mass in the central regions of the parent galaxy generates tidal forces, which are larger than those experienced by subhaloes in traditional {\it N}-body simulations. Increased tidal forces substantially alter circular velocity profiles for satellites with pericentres less than $50$~kpc. We show that these two effects are strong enough to bring the predictions of subhaloes from CDM simulations into agreement with the observed structure of MW dSphs, regardless of the details of the baryonic processes.
\end{abstract}

\section{Introduction}
\label{introduction}

According to the concordance cosmological model, a large fraction of the matter in the Universe is collisionless and dark. This Cold Dark Matter (CDM) model is supported by observations of large-scale structure as traced by galaxies and the emergence of this structure from initial density fluctuations, as observed in the cosmic microwave background \citep[e.g.,][]{Komatsu:2011in}. However, inconsistencies between a number of observations and the predictions of the CDM model on small (galactic) scales have understandably produced concern and attracted a great deal of attention from both observers and theorists.

The first of the small-scale problems for the CDM model is the `missing satellites problem' \citep{Klypin:1999ej, Moore:1999ja}. Numerical simulations of the CDM theory predict a large number of low-mass substructures to be present in a dark matter (DM) halo comparable in size to the Milky Way (MW). However, the number of subhaloes predicted is much larger than the number of satellite galaxies observed around the MW and in the Local Group (LG). This holds true even after accounting for the extreme low surface brightness dwarf galaxies that have been recently found using the Sloan Digital Sky Survey \citep{Koposov:2008ja, Koposov:2009kh, Tollerud:2008eq, Bullock:2010jz}.

The observed abundance of satellite galaxies can be reconciled within the context of CDM by invoking galaxy formation physics. Star formation (SF) can be suppressed in low-mass subhaloes by the photoionization of cold gas \citetext{\citealp*{Quinn:1996wd, Bullock:2000bn}; \citealp{Benson:2002ek, Somerville:2002ky}; \citealp*{Okamoto:2008ha, Okamoto:2009jd}; \citealp{Koposov:2009kh}} and in larger subhaloes through energetic supernovae (SNe) feedback \citep[e.g.,][]{Dekel:1986cv, Dekel:2003id, Governato:2007dq}. Thus, the complex physics of galaxy formation may solve the missing satellites problem. However, producing realistic simulations remains challenging as high numerical resolution is required to capture many relevant physical processes.

Differences between observed and predicted inner density profiles, termed the `core-cusp' issue, present another small-scale problem for the CDM theory. DM haloes are predicted to have a universal density profile ($\rho \sim r^{\alpha}$)  with a `cuspy' $\alpha \sim -1$ inner slope from the numerical simulations of \citet*[][NFW profile hereafter]{Navarro:1996ce,Navarro:1997if}. Subsequent numerical results find haloes with steep inner density slopes and NFW or Einasto profiles \citetext{\citealp*{Dekel:2003gu}; \citealp{Dekel:2003ew}; \citealp{Reed:2005fe,Diemand:2008hr,Springel:2008gd,Navarro:2010hn}}. In contrast, observations have repeatedly found constant density `cores,' where $\alpha \sim 0$ \citetext{\citealp{deBlok:2001hs, deBlok:2002bw, Swaters:2003kf, Strigari:2006ik, Walker:2009ez}; \citealp*{Strigari:2010kl}; \citealp{Oh:2011fk, Walker:2011eg, Hayashi:2012fn, Wolf:2012vc}}.

Numerical studies have examined how baryonic effects can alter the inner regions of galaxies. Repeated fluctuations of the central galactic potential due to SF or SNe feedback can transform a central cusp into a core \citetext{\citealp*{Mashchenko:2006ff, Mashchenko:2008fa}; \citealp{Governato:2010ed, Governato:2012cw, Pontzen:2012jg}}. This requires strongly coupled SNe feedback (i.e. $\Mvir = 10^{10}~\Msun$ with $M_{*} = 10^{7}~\Msun$ for a 1~kpc core) and, as discussed in \citet{Penarrubia:2012dy}, directly opposes the suppression of low-mass galaxy formation often invoked to solve the missing satellites problem. On the other hand, single blowout events can reduce the central densities but are unsuccessful in creating cores \citetext{\citealp{Gnedin:2002bu}; see however, \citealp*{Navarro:1996ws}}.

Another issue examines the abundance and structure of relatively massive satellites. Named the `too big to fail' problem, it questions if CDM predicts too many satellites with large circular velocities ($\vmax$) that are unable to host LG dwarf spheroidal galaxies (dSphs). CDM correctly predicts the abundance of {\it very} large satellites like the Magellanic Clouds \citep{Busha:2011jp, Busha:2011gk} and dwarf field galaxies with $\vmax>50~\kms$ \citep{Tikhonov:2009eh, TrujilloGomez:2011js}. However, the situation is different for satellites with $30~\kms < \vmax < 50~\kms$. Comparing the circular velocity profiles of subhaloes in the Aquarius {\it N}-body simulations with the circular velocities measured at the half-light radii for LG dSphs, \citet*{BoylanKolchin:2011ky, BoylanKolchin:2012id} find too many very compact, large subhaloes. For example, haloes with  $\Mvir \sim 1.4\times10^{12}~\Msun$ have $\sim$25 satellites with infall circular velocities $\vmax > 30~\kms$ within 300~kpc of the halo's centre \citep{BoylanKolchin:2012id}. A given halo contains at least seven satellites expected to host luminous galaxies, but without MW counterparts.

Within the context of the CDM model, there are a couple of solutions which may explain these inconsistencies. First, the CDM cosmology adopted for the Aquarius haloes uses $\sigma_8=0.9$ instead of the lower concordance value of $\sigma_8=0.8$. Subhaloes in lower $\sigma_8$ models have lower abundances and their density profiles are less concentrated, such that they are more susceptible to tidal effects \citep{DiCintio:2011df}. However, as discussed in \citet{BoylanKolchin:2011ky}, the `too big to fail' subhaloes are found when examining both the Via Lactea II simulation ($\sigma_8 = 0.74$) and the Aquarius simulations ($\sigma_8 = 0.9$). Furthermore, \citet{Wang:2012jg} find little change between the subhalo circular velocity functions using {\it Wilkinson Microwave Anisotropy Probe} ({\it WMAP}) and {\it WMAP}7 cosmologies.

Secondly, the abundance of substructures above a fixed circular velocity scales linearly with the host halo mass. Lowering the mass of the MW by $\sim 20$ per cent, from $\Mvir \sim 10^{12}$ to $\Mvir \le 8 \times 10^{11}~\Msun$, reduces the abundance of massive, dense subhaloes such that their absence around the MW can be attributed to a statistical fluke \citep{Wang:2012jg, VeraCiro:2013fy}. However, lowering the mass of the MW reduces the probability of hosting Magellanic Cloud sized galaxies \citep{BoylanKolchin:2010bd, Busha:2011gk}. Additionally, if Leo I is a bound satellite, the mass of the MW is at least $\Mvir = 1.6\times10^{12}~\Msun$ \citep{BoylanKolchin:2013et}. 

Another possibility is that baryonic processes, such as stellar feedback and photoionization, in combination with subsequent tidal stripping, impact the structure of massive subhaloes. Cosmological simulations including baryonic physics have produced mixed results on their ability to solve the `too big to fail' problem \citep{DiCintio:2011df, DiCintio:2013fo, Brooks:2012wa, Parry:2012fd, Zolotov:2012hi}. The generation of central cores in satellite density profiles due to baryonic physics has been found to alleviate the tensions between the observations and CDM predictions \citep{Brooks:2012wa, Zolotov:2012hi}. However, accurately capturing baryonic processes requires hydrodynamical simulations run with $\sim10$ pc resolution, which is extremely computationally expensive. Another complication is added when mimicking baryonic processes with various subgrid models, which can differ significantly in the included feedback mechanisms \citep{Ceverino:2009ke,Colin:2010ee}.

Finally, if observations and predictions of CDM do not agree after including realistic baryonic effects, the differences will provide a unique opportunity to probe the nature of DM. Alternative scenarios, such as warm or self-interacting DM, have been proposed to explain why small-scale CDM model predictions are unable to match observations \citetext{\citealp{Lovell:2012gp}; \citealp*{Vogelsberger:2012dy}; \citealp{Peter:2013iu, Rocha:2013bo}}.

In this paper, we investigate modifications to the CDM expectations induced by the presence of baryons.  We focus on well-established and undisputed facts about the behaviour of baryons in the MW, namely (i) the fraction of mass in baryons is substantially below the cosmic mean for subhaloes that host dSphs and (ii) there is a baryonic concentration at the centre of the MW halo.

Cosmological {\it N}-body simulations of large-scale structure are initialized taking into account the effect of baryons on matter fluctuations. However, while evolving these simulations forward in time, it is common to assume that {\it all} matter, including baryonic matter, is collisionless. This implies that the baryonic distribution follows the matter distribution in an unbiased fashion. This assumption works well when calculating large-scale properties of the mass distribution, but predictions must be interpreted with caution for smaller, halo sized scales.

The simplifying assumption has two consequences in the context of the `too big to fail' galaxies. Gas-poor objects, like dSphs, should have reduced baryonic mass and the central parts of the MW should have more baryonic mass, than assigned in simulations. In reality, baryons should be removed from dSphs and also would dissipate in a host galaxy to form a disc. This leads to a twofold effect, causing the dSphs to be more resistant to tidal stripping in conventional {\it N}-body simulations than in reality. This allows massive subhaloes to preserve their structure and high circular velocities and survive in simulations down to redshift zero.

The paper is organized as follows. We discuss different aspects of dSphs and possible scenarios for their formation in Section~\ref{sec:introdSphs}. In Section~\ref{sec:method}, we present the details of our simulations and analytical methods. The effects of baryon removal for isolated galaxies are studied in Section~\ref{sec:baryonremoval}. In Section~\ref{sec:tidalstripping}, we present results of tidal stripping. The discussion and conclusions are in Sections~\ref{sec:discussion} and \ref{sec:conclusions}, respectively.

\section{Formation and evolution of dwarf spheroidals}
\label{sec:introdSphs}
Dwarf spheroidals are DM-dominated galaxies with very little gas and a low surface brightness, old, spheroidal stellar component. Formation scenarios for dSphs are currently constrained by observations of LG dwarfs and hydrodynamical and {\it N}-body simulations. In order to address why dSphs have a lower baryonic fraction than the cosmic mean, we must examine their formation and evolution. This information will guide us when we set the parameters of dwarf galaxies in our simulations.

The radial distribution of dSphs around their central galaxies, their cumulative velocity function, their morphologies and their luminosity function may provide valuable information on their evolutionary history. For example, it is well known that 95~per cent of dSphs are located within 250~kpc of the centres of known groups \citetext{e.g., \citealp{Karachentsev:2005gb}; \citealp*{Karachentsev:2005jn}}. In other words, there are hardly any isolated dSphs. Even inside the LG, dSphs are preferentially found in the inner regions around the MW and M31. These results imply that there are unique environmental forces shaping the structure and baryon fraction of dSphs.

It has long been suggested that dSphs were once gas-rich dwarf irregular galaxies (dIrrs) that have subsequently lost their gas due to ram pressure stripping and consumption during SF \citep{Einasto:1974fh, Lin:1983ha, Kormendy:1985bl}. This theory has been examined in the context of several successful simulations.

\citet*{Kravtsov:2004he} proposed a realistic model for the formation of dSphs able to create luminous galaxies that match the spatial distribution and cumulative velocity function of observed LG satellites using self-consistent cosmological simulations and a SF model. The luminous dSphs began as initially massive isolated galaxies that underwent tidal stripping to reduce their mass and circular velocity. As dSphs progenitors inhabited more massive haloes before experiencing tidal stripping, this model explains how the LG dSphs with $\vmax \la 30~\kms$ were able to retain their gas, continue to form stars after reionization and remain luminous.

Both internal and environmental effects strongly influence the mass assembly history of a dwarf galaxy. This can result in dramatically different density profiles for galaxies of the same total mass. We use controlled numerical experiments to investigate different mass removal techniques with the same initial density profile. Cosmological simulations which show that halo density distribution follow the NFW profile assume a universal fraction of baryons. Thus, our `DM-only' simulations using an NFW profile include a cosmological fraction of baryons and the appropriate contraction due to the inclusion of this mass.

The use of the NFW profile for our initial conditions is motivated by the observations of gas-rich, dIrrs, which are likely progenitors of dSphs. At 1~kpc the fraction of baryons in gas-rich dwarfs is observed to range from one third to four times the cosmological baryon fraction, and on average it ranges from one to three times the cosmological baryon fraction, $\gamma_{\rm DM} \sim 0.43$--$0.82$ \citep[from seven dwarfs from The $\HI$ Nearby Galaxy Survey (THINGS) in figs 3 and 4 of][]{Oh:2011fk}. As the NFW profile rests on the assumption of a cosmological baryon fraction everywhere, this density profile already includes the adiabatic response of DM to the presence of a cosmological fraction of baryons.

Due to a lack of certainty in the galaxy formation physics in these haloes, in Sections~\ref{sec:method} and \ref{sec:baryonremoval} we test a variety of methods to remove the baryonic component from dSph progenitors. These include (i) removing the baryons before the halo forms, mimicking dSph formation in a baryon-poor environment, (ii) removing the baryons instantaneously to reproduce sudden gas removal, as when a dIrr loses gas due to ram pressure stripping as it falls into a host halo and (iii) removing the baryons slowly from the halo, in order to test how the speed of gas removal impacts the structure of the dwarf.

\section{Methodology}
\label{sec:method}

\subsection{Cosmological hydrodynamical simulation}
In order to obtain realistic estimates of how baryon removal impacts a satellite's mass distribution we ran cosmological hydrodynamical simulations. We simulated a $\Mvir(z=0) = 3\times10^{10}~\Msun$ isolated dwarf galaxy run with the same initial conditions twice. One simulation was run without gas removal and the other was run with all gas heated just before $z=1$.

To perform these simulations, we used the adaptive mesh refinement {\it N}-body+hydrodynamics code {\sc hydroart} \citep*{Kravtsov:1997iy,Kravtsov:1999vi}. The code is adaptive in both space and time, achieving higher resolution in regions of higher mass density. The following cosmological parameters were used: $\Omega_{\rm M} = 1~-~\Omega_{\rm \Lambda} = 0.3$, $\Omega_{\rm b} = 0.045$, $h=0.7$ and $\sigma_8 = 0.8$. The simulation zoomed in on a single halo inside a $10~h^{-1}$Mpc comoving box with a DM particle mass $m_{\rm p} \sim 10^5~\Msun$ and a maximum resolution of 60~pc at redshift zero.

The physical model used by the code includes many relevant processes, such as cooling from metals and molecules, a homogeneous ultraviolet (UV) background, metal advection and stellar mass-loss. SF occurred stochastically in regions where the gas density is above $6~\rm{atm~cm^{-3}}$ and temperature is below $10^4~{\rm K}$. Stars are formed with a constant efficiency, converting 30~per~cent of each cell's gas mass into a star particle representing a simple stellar population with a Chabrier initial mass function.

Our subgrid feedback model included both thermal energy from SNe and shocked stellar winds and radiation pressure from massive stars ($M > 8~\Msun$). Thermal energy was deposited in the star particle's cell at a constant rate, $\Gamma = 1.59\times 10^{34}~{\rm erg~s^{-1}}~\Msun^{-1}$, for 40~Myr without artificially delaying cooling. The radiation pressure was calculated using $P_{\rm rad} = (\tau_{\rm UV} + \tau_{\rm IR}) L /(c\, r^2)$, where $\tau_{\rm UV} = 1$ and $\tau_{\rm IR} =50$ are the optical depth of dust to UV and infrared (IR) photons, respectively, $L$ is the total luminosity of the star cluster \citep[calculated using {\sc starburst}99;][]{Leitherer:1999jt}, and $r$ is half the size of the star particle's cell. This pressure was added to the hydrodynamical equations for 3~Myr in the star particle's cell. Radiative feedback is implemented in a similar way as in \citet*{Hopkins:2011fk} and \citet{Agertz:2013il}.

More details on the implementation of the SF and feedback model can be found in \citet{TrujilloGomez:2013ts}. The isolated dwarf galaxies simulated using this new model are in very good agreement with observations. In particular, they have no bulges, slowly rising rotation curves, and constant or slowing increasing SF histories.

We created two isolated dwarf galaxies. The first run, the fiducial run, was performed as described above with no alterations. The second run, which we call the `no gas' simulation is identical to the fiducial run until just before $z=1$. At this time we removed the galaxy's gas to emulate the gas stripping that occurs as a satellite falls into its host halo. To achieve this, we artificially increased the high-resolution gas cell temperatures to $10^6$~K and halted their cooling until redshift zero. This removed the majority of the gas mass and prevented further SF. Any remaining bound gas existed as a hot diffuse medium.

These simulations do not include the effects of tidal stripping of DM and stars by the host potential. As such, they provide an estimate of the minimum effect of baryonic physics on satellite density profiles. In Section~\ref{sec:baryonremoval}, we use these simulations to test the adiabatic expansion of a halo due to baryonic losses.

\subsection{{\it N}-body simulations}
\label{nbody}
We performed controlled collisionless {\it N}-body simulations to quantify the gravitational effects of baryon removal from subhaloes and the enhanced tidal effects of a MW disc on subhaloes. In what follows, we describe the code and the initial conditions used. These simulations incorporated the undisputed net results of baryonic processes.

\subsubsection{Code and initial conditions}
\label{subsec:ICs}
We used a direct summation {\it N}-body code with the leapfrog time-stepping integration scheme, a constant time step $\Delta t = 6.4\times 10^5$~yr and a Plummer softening of $\epsilon = 20$~pc.  All resultant simulations retained an energy conservation of $\Delta E/E < 10^{-5}$. The code was run in parallel using both open multiprocessing and message passing interface libraries.

Initial conditions were set up by solving the equilibrium Jeans equation to obtain the rms velocity dispersion $v_{\rm rms} (r)$ as a function of radius for a dwarf with a spherical NFW density distribution given by
\begin{align}
  \rho_{\rm sat}(r) &= \frac{v^2_{\rm max}}{4 \uppi G r^2_s}\frac{x_{\rm max}}{\mu(x_{\rm max})}\frac{1}{x(1+x)^2},\quad   x\equiv r/\rs, \label{eq:NFWa}\\
  \mu(x) &= \ln(1+x)-\frac{x}{(1+x)}.  \label{eq:NFW}
\end{align}
Here $\vmax$ is the maximum circular velocity of the dwarf and $x_{\rm max}$ is the distance at which the velocity peaks, where $x_{\rm max} = 2.163$. 

The distribution was truncated at the outer radius $r_{\rm out}=16\rs$. This truncation radius is close to the expected virial radius for haloes with masses typical for large satellites of the MW. The exact value of $r_{\rm out}$ is unimportant as the satellite will be severely tidally stripped when it falls into the potential well of the MW.

When setting initial conditions, we used multiple masses for particles.  In the inner regions, the particle mass, $m_1$, was small and nearly constant. The mass increased with distance as
\begin{equation}
            m_1(r) = m_1(0)\left[1+\left(\frac{r}{\rs}\right)^2\right]\,,
\end{equation}
where $m_1(0)$ is a normalization constant.  This prescription allowed us to resolve the central region of the dwarf with small particles and at the same time have an extended halo.

Throughout the dwarf's evolution, particles of different mass moved in radius so that massive particles occasionally came to the centre. We monitored the situation and found that small particles always dominated the central $\sim 1$~kpc region with a very low contribution from more massive particles. For a 200~000 particle simulation run for 10 dynamical times, the average particle mass increased by just 2~per cent within the central 100 particles and increased by 6~per cent within the central 1000 particles. Additionally, simulations of both constant and variable mass particles, run for 16 dynamical times, show that variable mass particle runs accurately reproduce the results of constant mass particle simulations.

We assumed an isotropic Gaussian distribution of velocities, which is known to produce small non-equilibrium effects for the NFW distribution. We allowed the system to evolve in isolation for 10 dynamical times to settle and come into equilibrium. The non-equilibrium effects were small at all times, with the maximum circular velocity within 2~per cent of initial value, after 10 dynamical times. The largest effect, a decline in the central density, resulted in a 30~per cent change from the initial density within $\sim 100$~pc. These changes remain at a low level and do not impact the basic NFW density or circular velocity profile shapes. All results are compared to the evolved isolated configuration.

Our simulations consisted of 200~000 variable mass particles. This corresponds to a simulation with $N_{\rm eff}=1.3\times 10^6$ constant mass particles in order to maintain the same mass resolution. The gravitational force acting on satellites was modelled using frozen external potentials, which roughly approximated the mass distribution in our MW Galaxy \citetext{\citealp*{Diemand:2007hb}; \citealp{Cuesta:2008gl}; \citealp*{Diemer:2013hu}}.

\subsubsection{Scaling models}
\label{subsec:scaling}
Our simulated dwarfs were created to test the effects of baryon removal and tidal stripping for large satellites. The initial dwarf has the following properties: $\rs = 4.0~\kpc$ and $\vmax=62~\kms$. Although this is a large maximum circular velocity for a redshift zero classical dSphs, it is reasonable for a dSph progenitor. The \citet{Kravtsov:2004he} model shows that haloes with $50~\kms \la \vmax \la 80~\kms$ are transformed into present-day luminous dwarfs with $10~\kms \le \vmax \le 60~\kms$. Our initial dwarf parameters were chosen to mimic a  `too big to fail' \citep{BoylanKolchin:2012id} galaxy before infall. In agreement with the definition of `too big to fail' galaxies, the entire initial profile of our dwarf is inconsistent at $2\sigma$ with all LG dSphs. These initial conditions allow us to thoroughly test if the \citet{BoylanKolchin:2012id} results can be brought into agreement with observations by including baryonic effects.

Our host galaxy was matched to the Aquarius E halo virial mass with an added baryonic component. The disc component has parameters $r_{\rm disc} = 3~\kpc$ and $M_{\rm disc} = 6 \times 10^{10}~\Msun$ and an NFW halo has parameters $\rs = 25~\kpc$ and $\Mvir = 1.39 \times 10^{12}~\Msun$ ($\vcirc = 179~\kms$). These are reasonable values for the MW halo and disc mass, which are estimated to be $M_{\rm tot} = 0.7$--$3 \times 10^{12}~\Msun$ \citetext{\citealp*{Wilkinson:1999cc, Sakamoto:2003jw}; \citealp{Smith:2007fi, Li:2008ji, Xue:2008kb, Kallivayalil:2009ho, Guo:2010do}; \citealp*{Watkins:2010fc}} and $M_{\rm disc} \sim 5.5 \times 10^{10}~\Msun$ \citep{Gerhard:2002wx, Flynn:2006he}, respectively.

We calculated the circular velocity of the initial perturber at 8~kpc, to provide the reader with a comparison of the halo and the MW. The baryon values are $M_{\rm bar}(r < 8\; \kpc) = 4.5 \times 10^{10}~\Msun$ or a $\vcirc = 156~\kms$ and the DM values are $M_{\rm DM}(r < 8~\kpc) = 4.9 \times 10^{10}~\Msun$ or a $\vcirc = 162~\kms$. This produces a circular velocity of $\vcirc = 224~\kms$, which is close to the value of $\vcirc = 218 \pm 6~\kms$ at the solar neighbourhood, as found by \citet{Bovy:2012gj}.

When scaling {\it N}-body simulations to physical units, we chose the radius ($r_0$) and velocity ($v_0$) to be our free parameters, which left all other quantities, such as time-scale ($t_0$) and mass ($m_0$), dependent on the choice of radius and velocity scales.

\begin{table}
  \caption{Different dwarf galaxy cases. The initial and final maximum circular velocities during their isolated runs are listed.}
\label{tab:1}
\begin{tabular}{llccc}
\hline
Label & Description & $v_{\rm max, i}$ & $v_{\rm max, f}$ & $\frac{v_{\rm max, f}}{v_{\rm max, i}}$ \\
\hline
NR & No baryonic mass removal         & 62.1 & 64.2 & 1.03\\
PR & Pre-halo formation mass removal        & 64.2 & 57.6  & 0.90\\
ER & Exponential mass removal                    & 64.2 & 51.2  & 0.80\\
IR  & Instantaneous mass removal                & 64.2 & 51.0  & 0.79\\
\hline
\end{tabular}
\end{table}

\subsubsection{Setup of tidal stripping simulations}
\label{subsubsec:setuptidalstripping}
The simulations were run in the presence of an external MW-sized perturber after running in isolation. The no mass removal (NR) case was run with a pure NFW perturber, while all other cases were run with an NFW plus spherical exponential `disc' perturber. This setup allows the NR case to remain as the pure DM example, while the mass removal cases show the effects of including baryons (by adding a disc) in the simulation (see Table~\ref{tab:1} for naming conventions). Eight orbits were created for each satellite in order to compare the impact of orbital evolution on the structure of subhaloes.

All satellites were run in circular orbits at $r_{\rm orbit} = 50$, $70$, $100$ and $150~\kpc$ to examine the most extreme effects of tidal stripping from different distances. We also ran four elliptical orbits, 1:2e (150--70~kpc), 1:3e (150--50~kpc), 1:5e (150--30~kpc) and 1:4e (120--30~kpc), to provide more realistic results of tidal stripping. These orbits are less elliptical than the typical orbital ratio of 1:6 derived by \citet{Ghigna:1998ik}; however, they are well within the 90~per cent interval found by \citet{Diemand:2007hb}, which extends from $\sim$1:28 to $\sim$2:3 orbits. A description of each dwarf galaxy's orbits is provided in Table~\ref{tab:2}.

The pericentres of the elliptical orbits range from 30 to 70~kpc. These orbits were created to be realistic generalizations of LG dSph orbits as well as to investigate an interesting regime of tidal disruption. In our truncation radius estimates (Section~\ref{analytics}), we find that a satellite in the presence of a MW with a baryonic disc component should have interesting tidal evolution if it had a pericentric distance of $\sim 40$~kpc. This occurred when the truncation radius, $\rt$, had values from two to four times the scale radius, $\rs$. Our circular and elliptical orbits were in part chosen to examine a large range of truncation radii from extreme truncation to minimal tidal stripping.

Results from our elliptical orbits will in general reproduce the range of known orbits for the LG dSphs from radial and tangential velocity measurements. These orbits are expected to be 20--100~kpc for Carina \citep{Piatek:2003dd}, 40--89~kpc for Ursa Minor \citep{Piatek:2005ii}, 66--129~kpc for Sextans \citep*{Walker:2008cm}, 68--122~kpc for Sculptor \citep{Piatek:2006jz} and 118--152~kpc for Fornax \citep{Piatek:2007ht}. Our work does not investigate the orbital evolution of dwarfs such as Leo I and Leo II, which are most likely on their first infall and orbit at extreme distances \citep{Lepine:2011if,Sohn:2013do}.

Although none of our simulations examine orbits with pericentres lower than 30~kpc, the work of recent cosmological simulations find a significant fraction of satellite orbits with small pericentric radii. In the Via Lactea simulations \citep{Diemand:2007hb}, almost all galaxies within the `earliest forming substructure' sample had orbits that fell within 30~kpc of the host. In the smoothed particle hydrodynamics simulations run by \citet{Zolotov:2012hi}, over half of all the subhaloes had orbits with pericentres below 30~kpc (private communication).  Any satellites in orbits with pericentres smaller than 30~kpc will have more extreme orbital evolution than the results presented here.

\begin{table}
  \caption{Orbital parameters for each satellite galaxy. Included are their initial maximum circular velocities and final maximum circular velocities after 5~Gyr. Labels that end in `c' denote a circular orbit and those that end in `e' denote an elliptical orbit.}
\label{tab:2}
\begin{tabular}{llccc}
\hline
Label & Description & $v_{\rm max, i}$ & $v_{\rm max, f}$ &
$\frac{v_{\rm max, f}}{v_{\rm max, i}}$\\
\hline
NR150c    & NR, 150 kpc orbit               & 64.2  & 60.7 & 0.95\\
NR100c    & NR, 100 kpc orbit               & 64.2  & 57.2 & 0.89\\
NR70c      & NR, 70 kpc orbit                  & 64.2  & 51.6 & 0.80\\
NR50c      & NR, 50 kpc orbit                  & 64.2  & 43.0 & 0.67\\
NR1:2e     & NR, 150 to 70 kpc orbit     & 64.2  & 56.5 & 0.88\\
NR1:3e     & NR, 150 to 50 kpc orbit     & 64.2  & 53.8 & 0.84\\
NR1:5e     & NR, 150 to 30 kpc orbit     & 64.2  & 51.2 & 0.80\\
NR1:4e     & NR, 120 to 30 kpc orbit     & 64.2  & 46.9 & 0.73\\
&&&&\\
PR150c     & PR, 150 kpc orbit                & 57.6 & 53.6 & 0.93\\
PR100c     & PR, 100 kpc orbit                & 57.6 & 49.5 & 0.86\\
PR70c       & PR, 70 kpc orbit                   & 57.6 & 42.1 & 0.73\\
PR50c       & PR, 50 kpc orbit                   & 57.6 & 28.1 & 0.49\\
PR1:2e     & PR, 150 to 70 kpc orbit       & 57.6 & 48.5 & 0.84\\
PR1:3e     & PR, 150 to 50 kpc orbit       & 57.6 & 44.7 & 0.78\\
PR1:5e     & PR, 150 to 30 kpc orbit       & 57.6 & 40.4 & 0.70\\
PR1:4e     & PR, 120 to 30 kpc orbit       & 57.6 & 34.2 & 0.59\\
&&&&\\
ER150c    & ER, 150 kpc orbit                 & 51.2 & 46.0 & 0.90\\
ER100c    & ER, 100 kpc orbit                 & 51.2 & 40.5 & 0.79\\
ER70c      & ER, 70 kpc orbit                    & 51.2 & 30.0 & 0.59\\
ER50c      & ER, 50 kpc orbit                    & 51.2 & 10.7 & 0.21\\
ER1:2e     & ER, 150 to 70 kpc orbit       & 51.2 & 39.6 & 0.77\\
ER1:3e     & ER, 150 to 50 kpc orbit       & 51.2 & 34.8 & 0.68\\
ER1:5e     & ER, 150 to 30 kpc orbit       & 51.2 & 29.9 & 0.58\\
ER1:4e     & ER, 120 to 30 kpc orbit       & 51.2 & 22.5 & 0.44\\
&&&&\\
IR150c     & IR, 150 kpc orbit                   & 51.0 & 45.8 & 0.90\\
IR100c     & IR, 100 kpc orbit                   & 51.0 & 39.9 & 0.78\\
IR70c       & IR, 70 kpc orbit                      & 51.0 & 29.0 & 0.57\\
IR50c       & IR, 50 kpc orbit                      & 51.0 & 10.6 & 0.21\\
IR1:2e      & IR, 150 to 70 kpc orbit         & 51.0 & 39.0 & 0.76 \\
IR1:3e      & IR, 150 to 50 kpc orbit         & 51.0 & 34.0 & 0.67\\
IR1:5e      & IR, 150 to 30 kpc orbit         & 51.0 & 29.2 & 0.57\\
IR1:4e      & IR, 120 to 30 kpc orbit         & 51.0 & 21.4 & 0.42\\
\hline
\end{tabular}
\end{table}

\subsection{Adiabatic expansion}
\label{subsec:AE}
The removal of baryons from a galaxy results in smaller mass inside a given radius $r$. In turn, this reduces the force of gravity, causing the galaxy to expand.  This is a known and qualitatively well-understood physical process \citep{Zeldovich:1980ta, Blumenthal:1986ie, Gnedin:2004hc}. It is also observed in many cosmological simulations that include hydrodynamics and SF \citetext{\citealp*{Colin:2006fp, Gustafsson:2006fr}; \citealp{Tissera:2010ka, Gnedin:2011tf}}.

Here we use the simplest prescription of \citet{Blumenthal:1986ie} to make an analytical model for the expansion of a galaxy when it loses mass. It assumes circular orbits for DM particles and uses conservation of angular momentum to derive a relation between the initial mass, $M_\rmi(r_\rmi)$, and radius, $r_\rmi$, and the final mass, $M_\rmf(r_\rmf)$, and radius, $r_\rmf$. Note that the prescription does not use adiabatic invariants; it uses an integral of motion -- the angular momentum. Thus, its limitation is not that the process must be slow (adiabatic), it is the assumption of circular motions. An extension of the prescription, provided by \citet{Gnedin:2004hc}, makes a correction due to non-circular motions.  As \citet{Gnedin:2004hc} show, in the regime of small compression (or expansion) the difference between the two approximations is small.

It is convenient to rewrite the \citet{Blumenthal:1986ie} prescription using the circular velocity, $V_{\rm circ} = \sqrt{{\rm GM}(r)/r}$ of a mass shell, where $M(r)$ is the mass inside radius $r$.  If $\alpha(r_\rmi)$ is the fraction of mass that remains inside $r_\rmi$ after the removal of baryons, then the final radius, $r_\rmf$, and circular velocity, $V_\rmf$, are given by
\begin{align}
  r_\rmf &= \frac{r_\rmi}{\alpha(r_\rmi)}, \label{eq:Expa}\\
  V_{\rmf} (r_\rmf)  &= \alpha(r_\rmi)\,V_{\rmi}(r_\rmi) =\alpha(\alpha\, r_\rmf)\,V_{\rmi}(\alpha\, r_\rmf).
\label{eq:Exp}
\end{align}

Note that the factor of $\alpha$ in equation~(\ref{eq:Exp}) enters linearly in both the radius, $r_\rmi$, and circular velocity, $V_\rmi$. As a result, the effect of the expansion is rather large.  This should be contrasted with the case that lacks expansion when mass $M(r)$ decreases by a factor of $\alpha$ but the radius stays the same: $r_\rmf=r_\rmi$. In this case, the final circular velocity, $V_\rmf$, declines only by a factor of $\sqrt{\alpha}$, which is a small effect.

For the baryon removal case, where $\fb \approx 0.17$ \citep{Komatsu:2011in}, we expect that $\alpha \approx 1-\fb \approx 0.83$. If dSphs do not undergo sudden mass removal, due, for example, to formation in a low $\fb$ region, $V_\rmf$ will decline by $\sim 0.91$. However, if dSphs experience a reduction in their $\fb$ from a sudden removal of mass (e.g. ram pressure stripping, winds), they will undergo adiabatic expansion and  $V_\rmf$ will decline by $\sim 0.83$.

\subsection{A tidal stripping model}
\label{subsec:atidalstrippingmodel}
Simulations of the effects of tidal stripping can take substantial computing resources, so it is helpful to have a simple model that can reproduce some simulation results. The following model imitates tidal stripping by iteratively removing mass from a satellite. After tidal forces remove some fraction of mass the halo expands and additional particles may become unbound, giving rise to another round of mass-loss. This occurs due to the elongated elliptical orbits of DM particles. Thus, mass removal from peripheral regions, where the tidal force is strong, leads to a lack of mass in the inner regions over time. Due to this process, even the central region of a halo may be affected by tidal stripping.

A realization of the NFW distribution with 500~000 particles was made using the prescription described in Section~\ref{subsec:ICs}. The large number of low-mass particles in the central region of the dwarf allowed us to make detailed calculations of tidal stripping without numerical errors.

During each iteration unbound particles were removed according to their total energy. The potential energy term is calculated using the satellite potential including bound particles within the tidal radius and an external spherically symmetric tidal force. The initial tidal radius, $\rt$, was parametrized by its ratio to the characteristic radius, $\rs$, setting the constant tidal potential normalization throughout all iterations.

After all unbound particles were removed, the halo was adiabatically expanded. The ratio of the bound particle mass within each radius to the mass at the previous iteration provides the $\alpha$ parameter. The coordinates were increased and the velocities were decreased according to $\alpha$ and equations~(\ref{eq:Expa}) and~(\ref{eq:Exp}). Finally, the new tidal radius was found and we proceeded with the next round of unbound particle removal.

Just as in our {\it N}-body simulations, we found that tidal stripping continuously removed mass from the halo without convergence. Some fraction of mass was removed even after many iterations. However, the rate of the mass removal dramatically depended on the initial tidal radius. There was minimal mass-loss when the tidal radius initially exceeded $\rt =4\rs$, but more substantial stripping occurred at smaller tidal radii. After 10 to 30 iterations of removing unbound particles, this simple model agrees with our {\it N}-body simulations run for 5~Gyr (see Section~\ref{subsec:elliptorbits}).

\begin{figure*}
\includegraphics[trim= .8cm 1.5cm 1cm 1cm,clip,width=0.37\textwidth,angle=90]{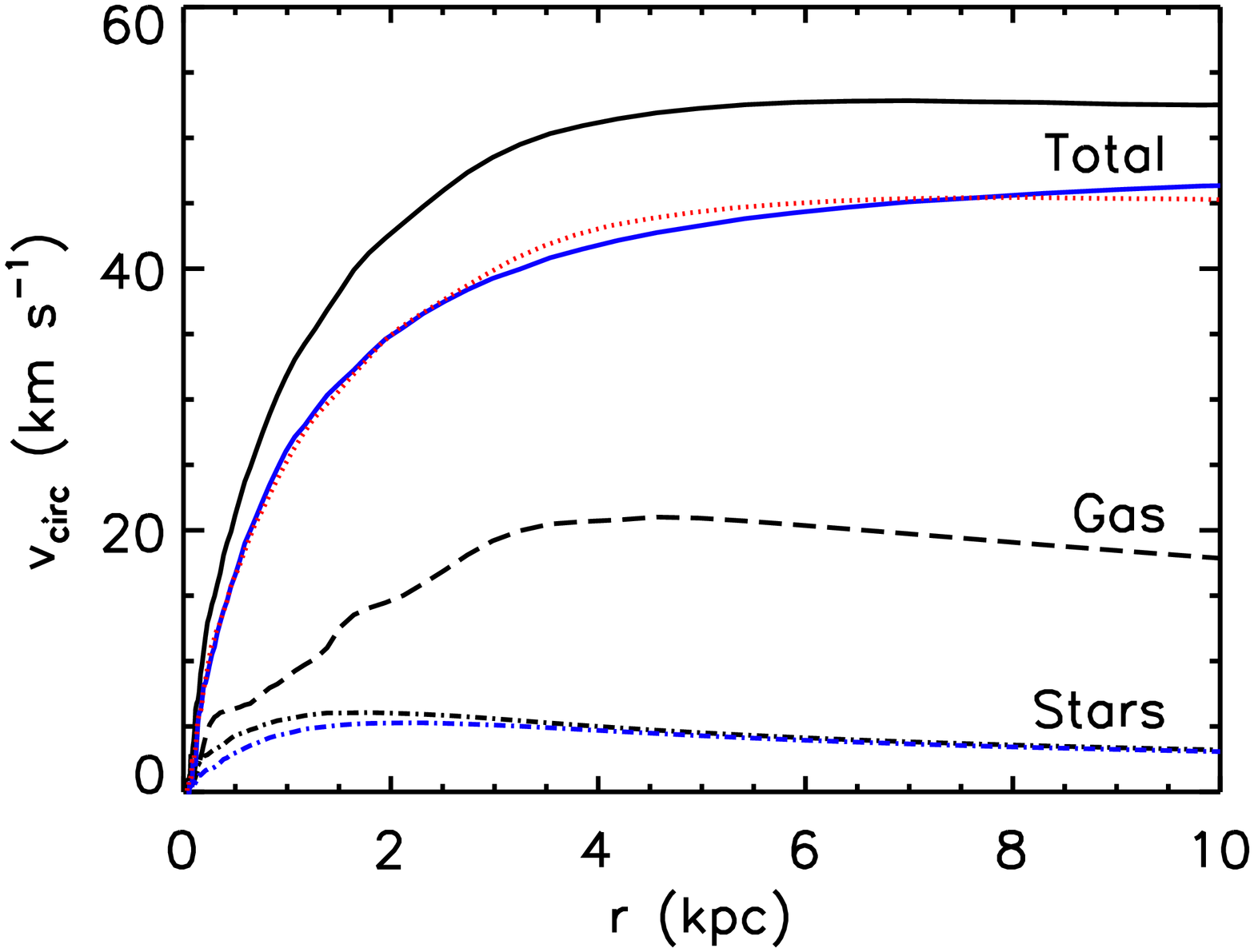}
\includegraphics[trim= .8cm 1.5cm 1cm 1cm,clip,width=0.37\textwidth,angle=90]{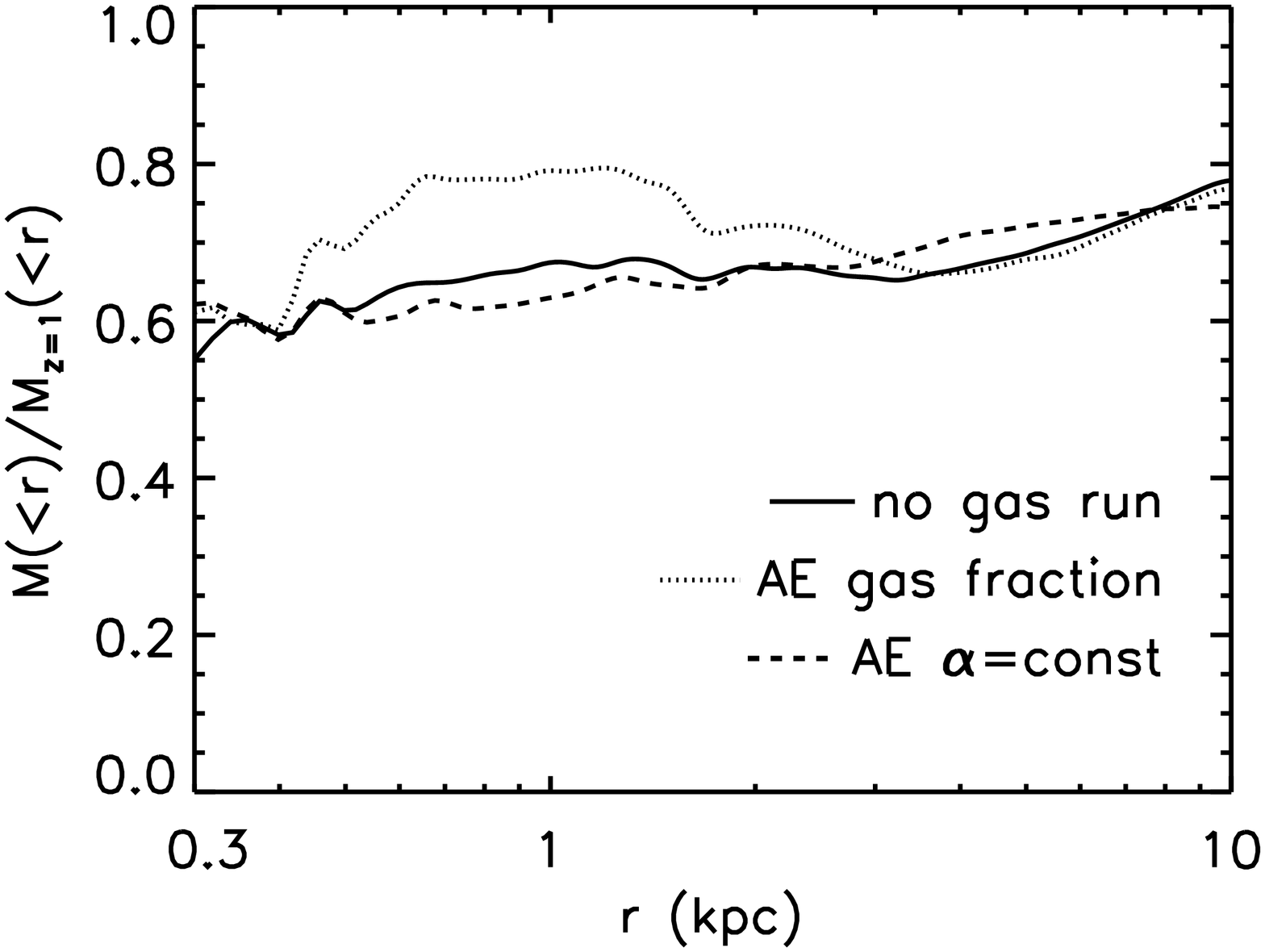}
\caption{
Results from the cosmological hydrodynamical simulations displaying the effects of gas removal from an isolated dwarf galaxy. Left: the circular velocity profile of the simulation at $z=1$, before the gas removal, and the `no gas' results at $z=0$. Solid lines represent the total circular velocity profile, the upper line (black) is at $z=1$ and the lower line (blue) is at $z=0$. The dotted (red) line is the adiabatic expansion model with constant removal. The bottom curves represent the circular velocity profiles of the gas (long dashed curve) and stars (dot--dashed curve) in the simulations at $z=1$. The $z=0$ stellar profile is shown as the lower long dashed curve (blue), and the gas is not shown as it has been removed. Right: here, the ratio of total mass within a given radius over the total mass within a given radius before gas removal (at $z=1$) is shown. The solid curve is the gas removal from simulation results at $z=0$. The adiabatic expansion models are shown as the short dashed curve for gas fraction removal and the dotted curve for constant 14~per~cent removal. The constant removal model predicts the expansion of the halo well, while the removal according to gas fraction underestimates the expansion in the inner 3~kpc.}
\label{fig:hydrosim}
\end{figure*}

\begin{table*}
\begin{minipage}{148mm}
   \caption{Mass values for the {\sc hydroart} simulated galaxies. Two isolated dwarfs were run, one fiducial model (left) and one where all gas within the galaxy was removed just before $z=1$ (right). The fiducial model increases its stellar mass and accretes more gas; however, the no gas dwarf retains only a hot diffuse gas component. The stellar mass in the no gas model is reduced from $z=1$ to $0$ due to stellar mass-loss, while the virial masses do not increase substantially from $z=1$ to $0$ for either model.}
   \label{tab:hydro}
   \begin{tabular}{lcccccc}
   \hline
   Model     &   \multicolumn{3}{c}{Fiducial} & \multicolumn{3}{c}{No gas}\\
                   & $\Mvir$ & $M_{\rm gas}(r<10~\kpc)$ & $M_{*}(r<10~\kpc)$ & $\Mvir$ & $M_{\rm gas}(r<10~\kpc)$ & $M_{*}(r<10~\kpc)$ \\
                   & (\Msun) & (\Msun) & (\Msun) & (\Msun) & (\Msun) & (\Msun) \\
   \hline
      $z=1$  &  $2.3\times10^{10}$ & $7.4\times10^8$ & $2.4\times10^7$ &  $ 2.3\times10^{10}$ & $ 2.5\times10^5$ & $2.4\times10^7$ \\
      $z=0$  &  $3.1\times10^{10}$ & $1.7\times10^{9}$ & $1.4\times10^{8}$  &  $ 2.5\times10^{10}$ & $2.2\times10^3$ & $2.2\times10^7$ \\
    \hline
\end{tabular}
\end{minipage}
\end{table*}

\section{Effects of baryon removal}
\label{sec:baryonremoval}

\subsection{Effects of baryon removal in hydrodynamical simulations: testing the adiabatic expansion model}
\label{hydro}

In this section, we examine the effects of sudden gas removal on the structure of a dwarf galaxy. We modelled this process in detail by performing two high-resolution cosmological simulations of the formation and evolution of a dwarf galaxy, including gas, SF and feedback. The hydrodynamical simulations presented here are meant to illustrate the general properties of an isolated dIrr. These results are not expected to reproduce properties of dSphs as there is no orbital evolution, but serve as a guide to initialize our tidal stripping simulations.

The fiducial run was completed with no removal of gas -- the system evolved without any intervention -- and mimics the evolution of a dIrr. As the left-hand panel of Fig.~\ref{fig:hydrosim} demonstrates, the simulation reproduces some generic properties of dIrrs: almost no bulge, a gently rising rotation curve and a larger gas mass than stellar mass.

In a second simulation, the `no gas' run, we ran the same simulation with the same physical prescriptions until just before $z=1$. At this point, all the gas was heated to $10^6$~K to mimic the sudden baryon removal that a dwarf would undergo upon falling into a host halo. Cooling was stopped in the heated cells, preventing further SF and allowing the galaxy to passively evolve until $z=0$. At redshift zero the `no gas' run retained a hot diffuse gas component of $\sim 2000~\Msun$ within 10~kpc and a total of $\sim 4\times10^5~\Msun$ within the virial radius. The general properties of the two dwarfs are listed in Table \ref{tab:hydro}.

At redshift one, the fiducial galaxy has the cosmological baryon fraction, $\fb = 0.17$, within the central 5~kpc, similar to observations of gas-rich dwarfs \citep{Oh:2011fk}. The inner 1~kpc region of the dwarf is slightly baryon deficient with a $\fb = 0.1$. The simulations presented are a realistic examination of mass-loss; however, a more dramatic change in structure would occur with a higher initial gas fraction in the central 1~kpc.

The stellar mass of the fiducial dwarf at redshift one is $\sim 2\times10^7~\Msun$ within 10~kpc, which corresponds to the stellar mass of Fornax, the most luminous of the classical dSphs. More typical dSphs have $M_* \sim 10^5~\Msun$. These galaxies should have SF shut off at earlier times or undergo lower SF rates than the simulation presented here to match observed dSphs. Star formation should be halted at infall time, which can occur earlier than redshift one. For instance, Draco and Ursa Minor stopped forming stars 10~Gyr ago \citep{Grebel:1998uz}, providing evidence for early accretion on to the MW. Therefore, prevention of the growth of stellar mass at redshift two and tidal stripping should be considered when comparing the stellar mass of this dwarf to the stellar mass of present-day dSphs. Nevertheless, the purpose of this paper is to address the issue of baryonic effects on the structure of massive subhaloes (which can have large stellar masses), not to investigate the origin of all of the dSphs.

Fig.~\ref{fig:hydrosim} compares the fiducial model to the `no gas' model at $z=0$, after the gas has been removed to allow the halo to adjust to the mass removal. In what follows, we use our adiabatic expansion model to examine how gas removal impacts a halo.

The left-hand panel of Fig.~\ref{fig:hydrosim} shows the circular velocity profiles of the gas, stars, and total mass for the simulations and the adiabatic expansion model. The mass removed was small, $M_{\rm gas}/M_{\rm tot} (< 5~\kpc) = 0.16$. However, removal of gas clearly resulted in halo expansion and a decline in circular velocity at every radius. Even in the central 1~kpc, where the gas fraction was lowest $M_{\rm gas}/M_{\rm tot} (< 1~\kpc) = 0.08$, the final expansion was substantial. The halo experienced a large reduction in mass within this radius, $M_{z=0}/M_{z=1} (< 1~\kpc)  = 0.67$.

The results of our cosmological hydrodynamical simulations show that halo expansion takes place when a fraction of mass is removed from the central region. The right-hand panel of Fig.~\ref{fig:hydrosim} shows the enclosed mass profile of the galaxy after gas removal divided by the fiducial model at $z=1$. In order to reproduce the results from the `no gas' run, we approximated the removal of the baryonic component in two ways. In our first trial, we removed the fraction of gas mass within a given radius from the dwarf's total mass. In our second trial, we removed a constant mass fraction of 14~per cent at all radii, which corresponded to the average baryonic mass lost within $\sim2$~kpc. Both models implemented the \citet{Blumenthal:1986ie} expansion, but used different mass removal at each radius.

The right-hand panel of Fig.~\ref{fig:hydrosim} compares the differences between the two expansion models. The removal of the enclosed gas fraction at each radius, where $\alpha(r) = 1 - f_{\rm gas}(<r)$, underestimates the expansion of the halo at all radii between 700~pc and 3.5~kpc. The underestimated expansion of this approach agrees with the \citet{Gnedin:2004hc} model who found that, after accounting for non-circular motions, expansion corresponds to more mass removed within a given radius than expected. The constant removal case, $\alpha(r) = 0.86$, gives results within 7~per~cent of the true expansion within the galaxy. This model is a straightforward technique to implement and produces fairly accurate results.

We consider a simple but realistic model in which baryonic effects cause the initial halo to retain a constant fraction of its density at every radius: $\alpha =\alpha_0=$~const. We use this analytic model to investigate the effects of varying amounts of mass-loss. Any halo profile can be written in the general form
\begin{equation}
   \rho(r) = \rho_0\Psi\left(\frac{r}{r_0}\right),
\label{eq:shape}
\end{equation}
where $r_0$ is a scale radius and $\rho_0$ defines the overall normalization of the profile. Here, the dimensionless function $\Psi(x)$ describes the density profile. For example, the NFW profile (equations~\ref{eq:NFWa} and~\ref{eq:NFW}) has the parameters $r_0 =\rs$ and $\Psi = 1/x(1+x)^2$.

When the halo loses $1-\alpha_0$ of its mass, it expands and changes the parameters $r_0$ and $\rho_0$, but the function $\Psi$ is preserved. Therefore, adiabatic expansion of a halo will preserve its shape but modify its parameters. Using equations~(\ref{eq:Expa})--(\ref{eq:shape}) and labelling variables with the subscripts $\rmi$ for initial and $\rmf$ for final parameters, we get
\begin{equation}
  \rho_{0,\rmf} =\alpha^4_0\,\rho_{0,\rmi}, \quad r_{0,\rmf} = \frac{r_{0,\rmi}}{\alpha_0}.
\label{eq:Exparameters}
\end{equation}
The equation for density normalization has a large power of $\alpha_0$. This means that the overall density normalization declines substantially with mass-loss. If we assume that most of the baryons were removed when a dIrr transforms into a dSph, then $\alpha_0\approx 1-\fb \approx 0.8$, which leads to a drop in density of $\rho_{0,\rmf}/\rho_{0,\rmi} \approx 0.4$.

Next, we pose the question of how much the density changes at a given physical radius after baryon removal. The answer depends on the particular form of the density profile:
\begin{equation}
  \frac{\rho_\rmf(r)}{\rho_\rmi(r)} = \alpha^4_0\,\frac{\Psi\left(\frac{r}{r_{0,\rmf}}\right)}{\Psi\left(\frac{r}{\alpha_0r_{0,\rmf}}\right)}.
\label{eq:Expar}
\end{equation}
If the function $\Psi$ is the power law $\Psi\propto x^{-\beta}$, then
\begin{equation}
  \frac{\rho_\rmf(r)}{\rho_\rmi(r)} = \alpha_0^{4-\beta}.
\end{equation}
For the NFW profile, the central cusp has a slope of $\beta=1$ so the density at a fixed radius declines by large factor, $\rho_\rmf/\rho_\rmi= \alpha_0^3\approx 0.5$, again assuming that $\alpha_0\approx 0.8$. The change in the outer region is much smaller as $\beta = 3$, so $\rho_\rmf/\rho_\rmi= \alpha_0\approx 0.8$. It is interesting to note that a density profile with a central core ($\beta =0$) experiences the largest decline in density, $\rho_\rmf/\rho_\rmi=\alpha_0^4$.

\citet{Gnedin:2002bu} find that sudden baryon removal does not impact the slope of the density profile, in agreement with this work. However, unlike their work our analytic model does not assume a particular form of a density profile, a halo spin parameter, or size of a baryon component removed. Therefore, our results are more general and allow us to predict the decline in density of any power-law profile to be dependent on $\rho_\rmf/\rho_\rmi= \alpha_0^{4-\beta}$.

\begin{figure*}
\includegraphics[trim= .8cm .5cm .8cm 1cm,clip,width=0.35\textwidth,angle=90]{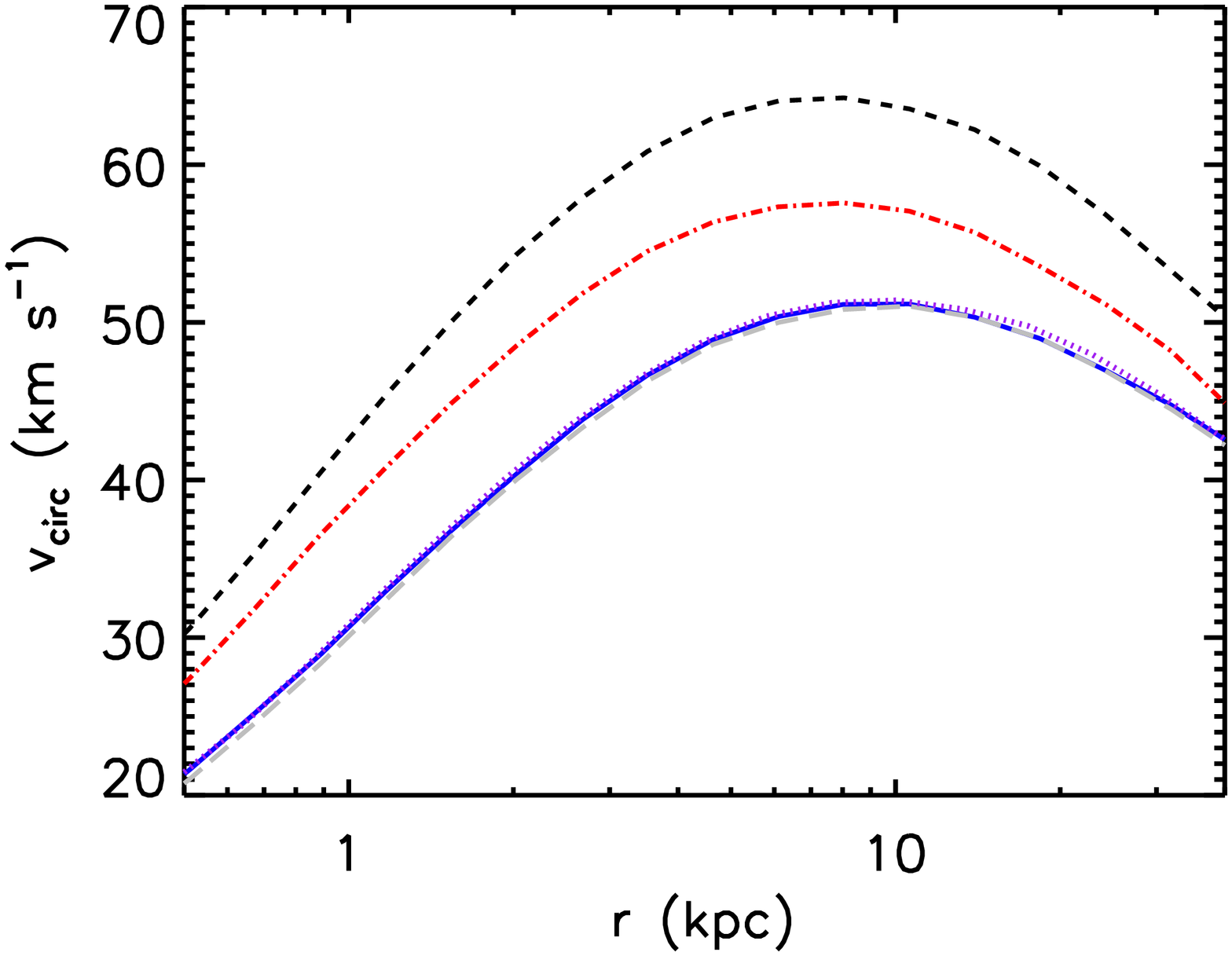}
\includegraphics[trim= .8cm .5cm .8cm 1cm,clip,width=0.35\textwidth,angle=90]{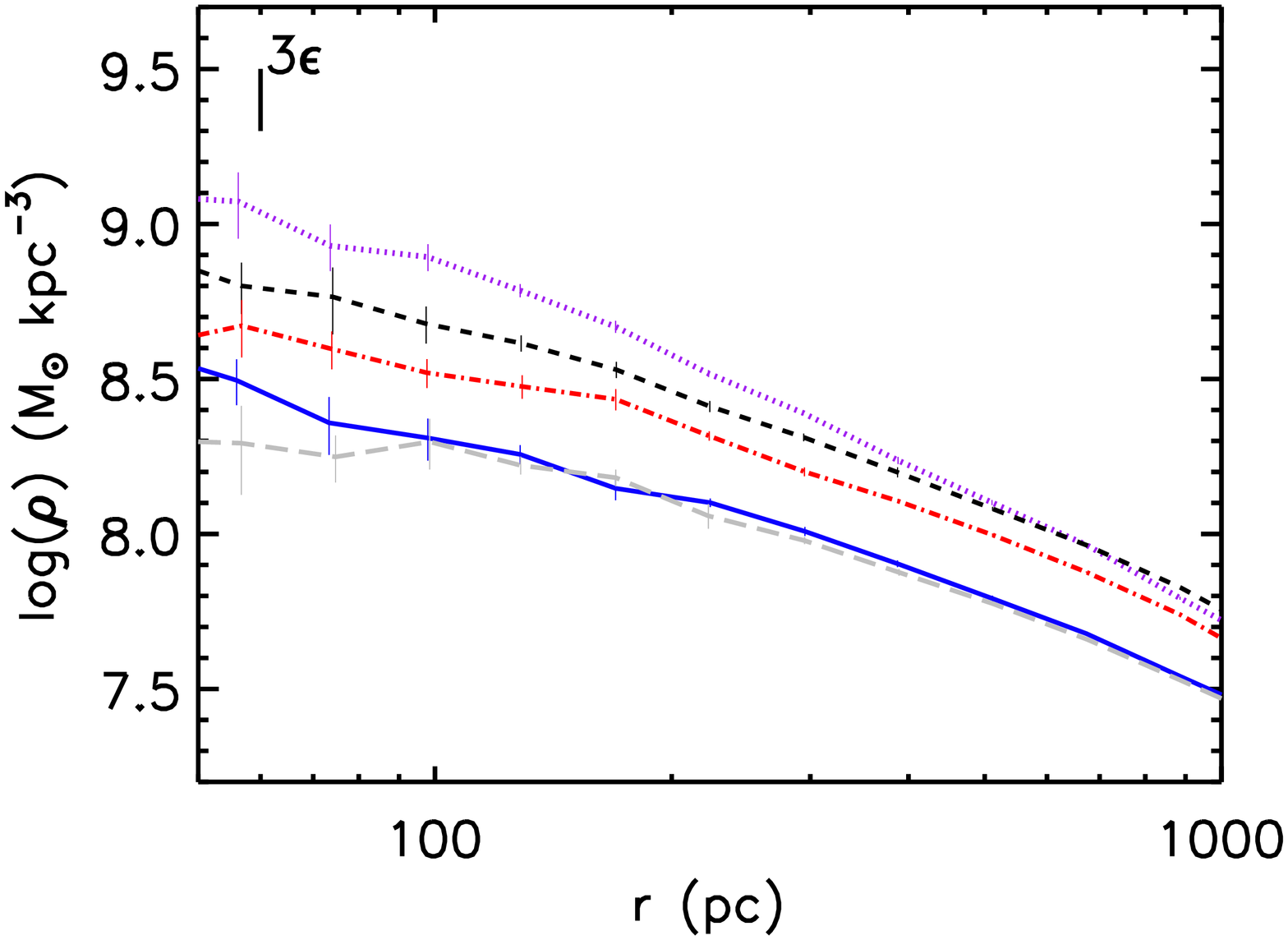}
\includegraphics[trim= .8cm .5cm .8cm 1cm,clip,width=0.35\textwidth,angle=90]{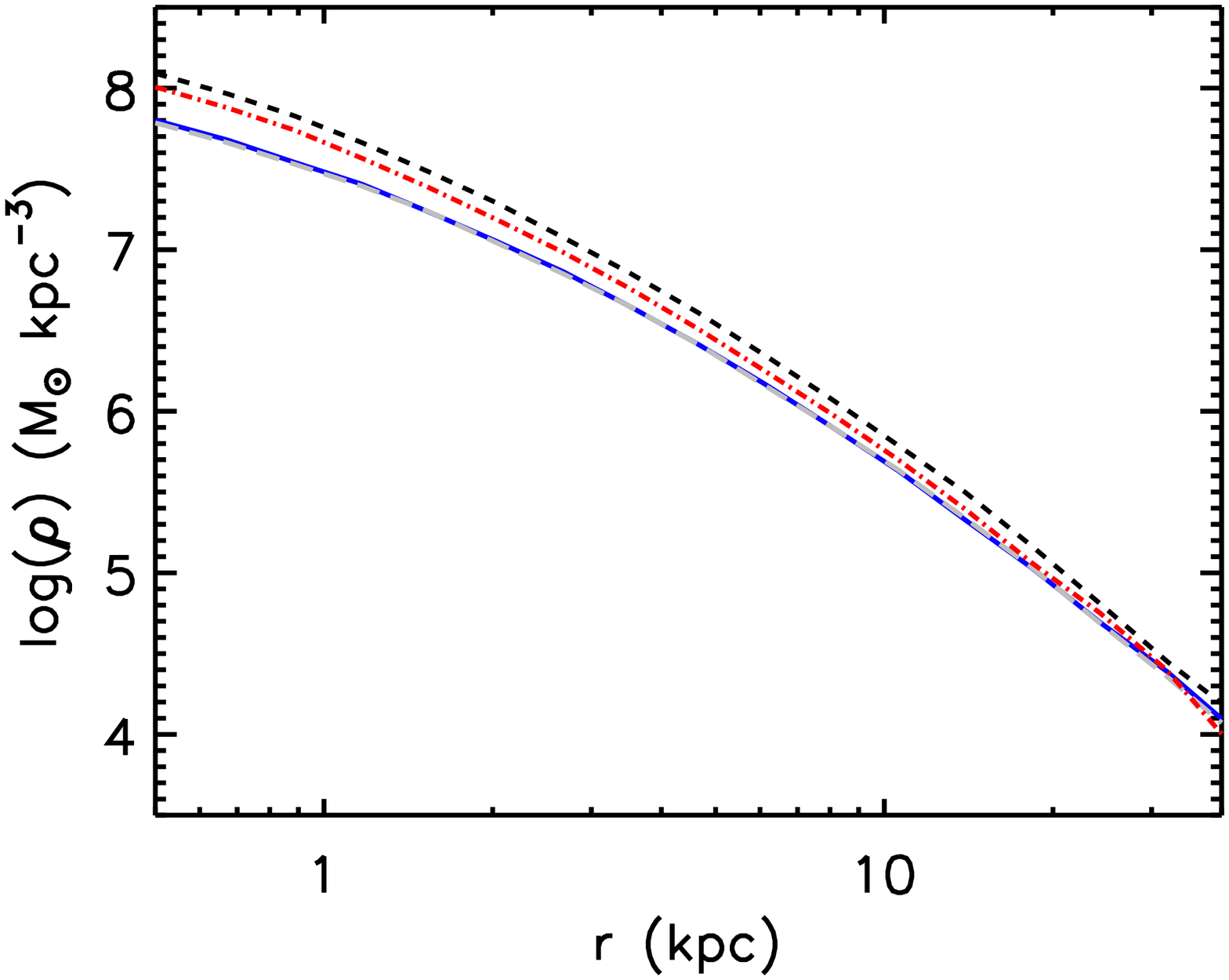}
\includegraphics[trim= .8cm .5cm .8cm 1cm,clip,width=0.35\textwidth,angle=90]{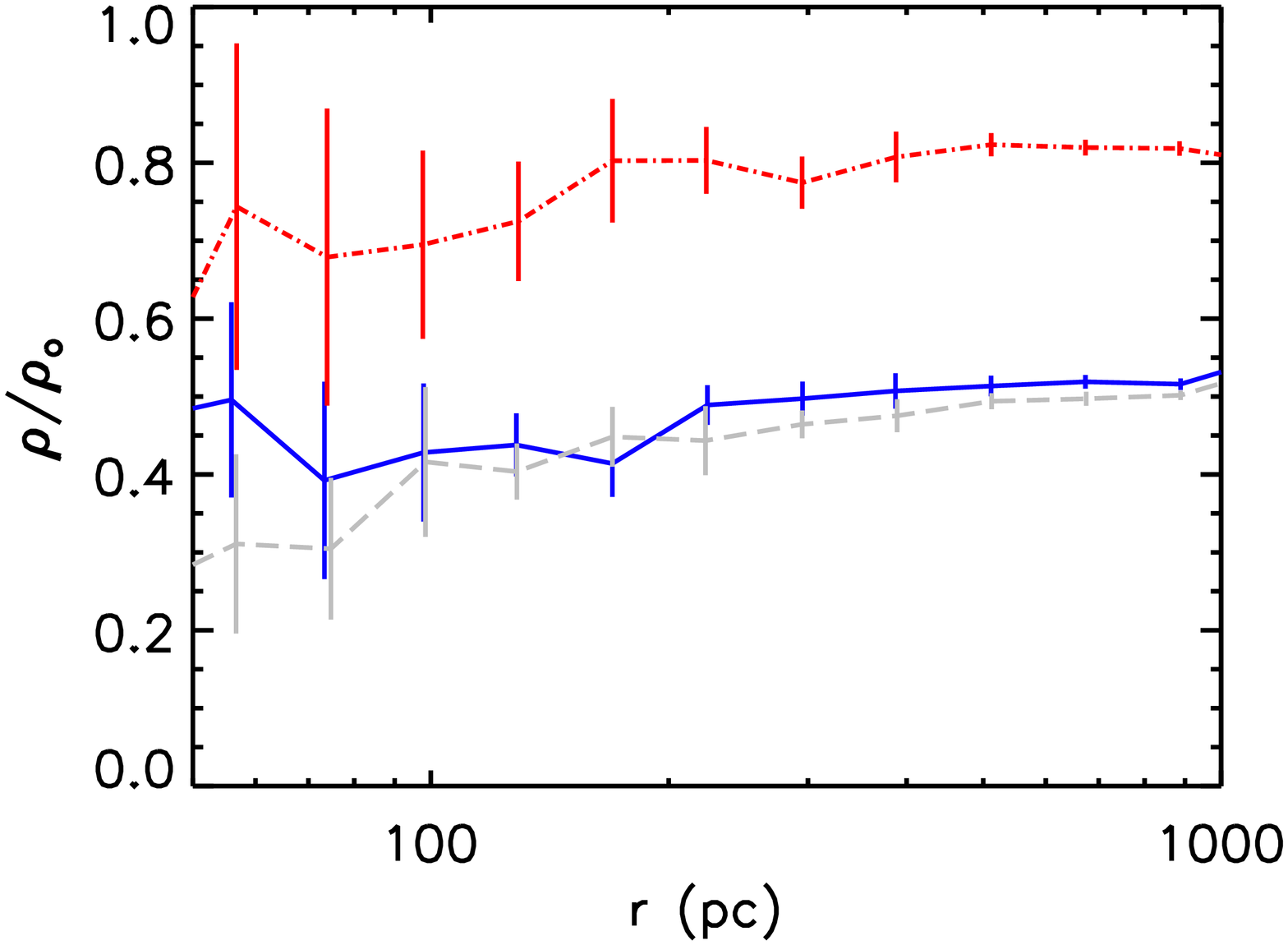}
\caption{Profiles of dwarf galaxies at their final moment in isolated evolution. In all plots, the short dashed (black) line shows the NR case, the dot--dashed (red) line shows the mass removal before the halo formed, the pre-halo formation mass removal (PR) case, the solid (blue) line shows the exponential mass removal (ER) case and the long dashed (grey) line shows the instantaneous mass removal (IR) case. Top Left: the circular velocity profiles of the four dwarf galaxies. An additional line, dotted (purple), displays the adiabatic expansion model that the NR halo would incur if it lost 20~per cent of its mass. It lies on top of the ER and IR profiles, indicating that our adiabatic expansion model accurately describes the mass removal process. Bottom Left: the density profiles of all four dwarfs over the same range as above. Top Right: the density profiles of all four dwarfs, zoomed into the central 50~pc to 1~kpc region. The initial state of the dwarf, before any isolated evolution or mass removal, is included as the dotted (purple) line. The NR halo begins to deviate from this profile at 500~pc. For reference, a radius of $3\epsilon$ is denoted on the plot. Bottom Right: the fractional deviation in density of the mass removal cases divided by the NR profile. The right two panels display averaged profiles in order to reduce shot noise. The errors displayed are the associated errors in the mean.}
\label{fig:equilib}
\end{figure*}

\subsection{Baryon removal in {\it N}-body simulations}
\label{subsec:baryonremoval}
Dwarf spheroidals are known for their high mass to light ratios, producing galaxies that are gas deficient with lower than average baryon fractions. To mimic this lack of gas, motivated by the results of previous section, we removed 20~per~cent of the mass from our dwarf to reflect the amount of baryonic matter lost when accounting for the lack of gas in dSphs. As the details of the removal process are uncertain, we created three mass removal cases to compare against a dwarf from pure DM simulations.

To test a few cases, we have run four dwarfs in isolation under different conditions (see Table 1 for naming conventions). Our first dwarf (no mass removal -- NR) was run for 1~Gyr from an equilibrium setup and is used as a benchmark for our comparisons to the other dwarfs which experience baryonic effects. This dwarf is supposed to mimic the results of collisionless simulations, which completely ignore baryonic effects. The three remaining dwarfs were modified using techniques that test a variety of ways a dwarf galaxy ends up with a small baryonic component.

On one extreme is the case where majority of baryons have never entered the DM halo. This model represents a dSph that formed in a region with low baryon fraction and subsequently does not contain a large gas fraction. Here, the dwarf will always stay in dynamical equilibrium with the correct baryon fraction. There is no expulsion of baryons and hence no change in the density profile due to the dynamics of baryons. The inner density profile of such a dwarf is large and it affects the efficiency of mass removal via tidal stripping. We label this scenario for the dwarf as pre-halo formation mass removal (PR). To distinguish this case from NR, this dwarf was evolved in isolation with an initial mass at 80 per cent of the NR case (i.e. $\alpha=1 - f_{\rm b} \sim 0.8$) and remained in equilibrium.

However, it is also possible that the dwarf once contained a baryonic component that was either expelled via SNe feedback, galactic winds and radiation pressure or removed by ram pressure and tidal stripping, similar to some of the processes which we explored in our hydrodynamical simulations. There is uncertainty in the exact mechanisms that cause this removal and whether such mechanisms are energetic enough to remove baryons from the galaxies. We follow the results of our tests with the hydrodynamical simulation and assume that the halo response is similar to a simple adiabatic expansion resulting from a constant fraction of total mass being removed from the halo. Since the time-scales for the baryon removal are also uncertain, we examine two different cases: one dwarf (instantaneous mass removal -- IR) began as the NR case and then had 20~per~cent of its mass instantaneously removed before evolving in isolation. Another dwarf (exponential mass removal -- ER) experienced a slow mass removal of 20~per~cent throughout its isolated evolution.

Details of the baryonic mass removal for the ER case is as follows. The mass began at the no removal level and then was reduced via an exponential function of the form: $m_\rmi(t) = m_{\rmi,0} (B + A e^{-t/t_0})$. Each particle began with a mass of $m_{\rmi,0}$ and currently has a mass of $m_\rmi(t)$. We have implemented $t_0 \sim 200$~Myr with $B = 0.8$ and $A = 0.2$. This equation allowed 20~per cent of the mass of each particle to be removed over an exponential time-scale, $t_0$. The larger the value of the time-scale, the slower mass is removed. The removal time-scale implemented is an order of magnitude larger than the dynamical time of the central region (e.g. $t_{\rm dyn}\sim 23$~Myr for $\vcirc(R = 1~\kpc) = 43~\kms$ and $t_{\rm dyn}\sim 16$~Myr for $\vcirc(R=0.5~\kpc) = 30~\kms$).

Note that we begin with an NFW density profile as the initial condition for all of our dwarfs. As discussed in Section~\ref{sec:introdSphs}, the NFW mass profile includes baryons which trace the DM in an unbiased manner. This initial condition thus already accounts for the adiabatic contraction of DM in response to the inclusion of the cosmological fraction of baryons. Observations of gas-rich dwarf galaxies (which are likely progenitors of dSphs) in the THINGS galaxy sample find average baryon fractions that range from one to three times the cosmological baryon fraction \citep{Oh:2011fk}. The NFW profile is therefore a physically motivated initial condition for our dwarfs.

In summary, our controlled numerical experiments examine the effects of baryonic mass-loss that dwarf galaxies undergo once they become satellites. In this manner, the three cases we investigate (PR, ER and IR) bracket the different physical scenarios that lead to the small baryon fraction in dSphs.

\subsubsection{Effects of fast and slow loss of baryons}
\label{subsubsec:baryonremoval}
The change in maximum circular velocity, $\vmax$, after each dwarf galaxy's isolated evolution is given in Table~\ref{tab:1}. The NR case, lists the $\vmax$ from the initial conditions and after 10 dynamical times of isolated evolution as the initial and final values, respectively. All results are compared to the final isolated NR $\vmax$ value, which represents the preferred equilibrium state of the dwarf galaxy.

Fig.~\ref{fig:equilib} shows the density and circular velocity profiles of the dwarfs at their final moment of isolated evolution. We see that the removal of baryons after the halo has formed, via instantaneous mass removal (IR) or exponential mass removal (ER), dramatically reduced the circular velocity and density profiles of the dwarfs without changing their shape. The haloes have responded to mass-loss by adiabatically expanding. Our adiabatic expansion model, explained in Section~\ref{subsec:AE}, matches the circular velocity profile of the IR and ER mass removal dwarfs using an $\alpha = 0.8$ to reduce $\vmax$ and increase $\rmax$.

In contrast, the removal of baryons before the halo has formed, as in the pre-halo formation mass removal (PR) case, caused $\vmax$ to decrease without an increase in $\rmax$. The baryons were not expelled from the halo but prevented from infalling so this halo did not undergo adiabatic expansion. Therefore, this halo has a larger density normalization, as discussed in Section~\ref{subsec:AE}.

The density profiles of the four dwarfs are most different within the central 1~kpc. To better examine the central 50~pc to 1~kpc region of the dwarfs, the right two panels in Fig.~\ref{fig:equilib} display the density profiles in the innermost regions. Comparing the inner density profile of the initial NFW halo and the evolved NR density profile they only differ within the central 500~pc.

Fig.~\ref{fig:equilib} also shows the three mass removal cases normalized by the NR density profile.
As predicted by Section~\ref{hydro}, the IR and ER mass removal prescriptions undergo adiabatic expansion and produce $\rho/\rho_0 \sim 0.5$. The PR halo has equilibrium mass removal so $\rho/\rho_0 \sim 0.8$, as $\alpha$ reduces the mass at every radius but the halo does not expand. The shape of all density profiles after mass removal did not change. No cores were created with or without adiabatic expansion, in agreement with \citet{Gnedin:2002bu}.

Since the expansion of a given DM halo will depend on the mass fraction removed from the halo, the cases we have explored examine the maximum and minimum effects of baryon removal. We tested the minimum removal case, where the halo forms baryon deficient (PR), and the maximum cases, where the halo suddenly removes 20~per~cent of its mass (ER and IR) and adjusts adiabatically. If a particular halo has a low gas fraction it will experience less dramatic adiabatic expansion. In the case of our hydrodynamical simulations, the halo is approximated to adiabatically expanded with a constant $\alpha = 0.84$, which would produce a halo between the PR and ER profiles, with $\rho/\rho_0 \sim 0.6$ in the inner region. We expect that the IR and ER cases will be more susceptible to tidal stripping than the PR case as they have experienced an increase in $\rmax$.

\section{Effects of tidal stripping}
\label{sec:tidalstripping}
\subsection{Truncation radius: analytical estimates}
\label{analytics}
In order to examine how the concentration of baryons in the inner part of the MW halo impacts a subhalo, we begin by presenting analytical estimates of the subhalo's tidal radius, depending on its mass, the main halo mass, its distance from the main halo and the inclusion or absence of baryons in the centre of the primary. One needs to be aware that the fate of the dwarf strongly depends on the tidal radius, $\rt$, to which it is stripped. We later show that if $\rt>4\rs$, there is no significant evolution of $\vmax$. However, tidal stripping can even alter the very centre of the dwarf when $\rt<2\rs\sim \rmax$. So there is a transition at $\rt \sim 2\rs$ with a small effect for larger $\rt$ and drastic stripping at smaller tidal radii.

If the subhalo has a mass profile of $m(r)$, which orbits a halo with mass profile of $M(R)$ at a radius $R$, then the tidal radius for the subhalo, $r_{\rm tide}$, is defined as the radius of the subhalo at which the gravitational force on a test particle from the subhalo equals the difference of the gravitational force (tidal force) from the halo on the subhalo and the test particle \cite[see e.g.][]{Klypin:1999bk},
\begin{align}
G\frac{m(r_{\rm tide})}{r_{\rm tide}^2} &= -\frac{\partialup}{\partialup R}\left[\frac{GM(R)}{R^2}\right] r_{\rm tide} \,. \\
\frac{m(r_{\rm tide})}{r_{\rm tide}^3} &=  \frac{M(R)}{R^3}\left[2-\frac{R}{M}\frac{\partialup M}{\partialup R} \right] \,.
\label{eq:truncrad}
\end{align}
The subhalo can also be stripped to an even smaller radius defined by resonances between the gravitational force from the subhalo and the tidal force from the halo. This resonant radius can be obtained by solving the following equation: 
\begin{equation}
\frac{m(r_{\rm res})}{r_{\rm res}^3}=\frac{M(R)}{R^3}\,.
\label{eq:truncrad_res}
\end{equation}
The smaller of $r_{\rm tide}$ and $r_{\rm res}$ is taken as the truncation radius. This is denoted by $\rt$ and defined as the radius to which the subhalo is stripped.

\begin{figure}
  \includegraphics[trim= .8cm 2cm 1.5cm 2cm,clip,width=0.35\textwidth,angle=90]{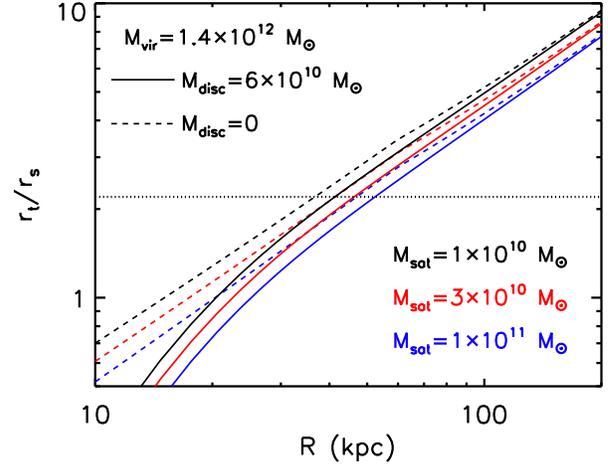}
  \caption{Dependence of the satellite truncation radius (in units of the scale radius) on the radial distance from the centre of its host halo. Satellites of varying masses ($M_{\rm sat} = 10^{10}$, $3\times10^{10}$ and $10^{11}~\Msun$) are shown from top to bottom in black, red and blue, respectively. Dashed lines show the truncation radii for satellites orbiting in the presence of a pure NFW halo ($M_{\rm vir}=1.4\times10^{12}~\Msun$) as contrasted with solid lines of the same colour showing a primary halo that also contains baryons ($M_{\rm disc}=6\times 10^{10}~\Msun$) that coalesced at the centre of the halo. The horizontal dotted line denotes $\rt = 2.2 \rs$. The internal structure of subhaloes can be dramatically affected when their orbits include pericentric passages with truncation radii below $2.2 \rs$.}
\label{figtrunc}
\end{figure}

For haloes following the NFW density profile, the mass profile $m(r)$ is given by
\begin{equation}
m(r) = \Mvir \frac{\mu(x)}{\mu(c_{\rm vir})}\,, \quad
            x \equiv \frac{r}{\rs},
\label{eq:subhalo_prof}
\end{equation}
where  $\rs$ is the scale radius of the halo and is related to the virial radius of the halo through the concentration parameter, $c_{\rm vir}=r_{\rm vir}/\rs$, and the function $\mu(x)$ is given by equation~(\ref{eq:NFW}). We assume that the subhalo mass profile follows the above distribution.

The main halo includes a DM component that follows the NFW profile and a baryonic component that is concentrated at the centre of the halo and is well inside the orbit of the satellite.  Therefore, the total mass profile of the halo is given by
\begin{equation}
M(R) = M_{\rm bar} + \Mvir(1-\fb) \frac{\mu(x)}{\mu(C_{\rm vir})}\,,
\label{eq:halo_prof}
\end{equation}
where $x=R/R_{\rm vir}$ and $C_{\rm vir}$ is the concentration of the halo. The parameter $\fb$ is the fraction of the virial mass that is in the baryonic component: $\fb = \Omega_{\rm bar}/\Omega_{\rm matter}$.  Here, $\Omega_{\rm bar}$ and $\Omega_{\rm matter}$ are the contributions of the baryons and the total matter as compared to the critical density of the Universe, respectively.  We assume that the baryonic mass of the halo is $M_{\rm bar}=6\times10^{10}~\Msun$, consistent with the baryonic mass estimates for the MW \citep*[e.g.,][]{Dehnen:1998gx,Klypin:2002bm}.
 
To determine the truncation radius, $\rt$, we substitute the mass profiles for the subhalo (equation~\ref{eq:subhalo_prof}) and the main halo (equation~\ref{eq:halo_prof}) into the tidal and resonant radii equations~(\ref{eq:truncrad}) and~(\ref{eq:truncrad_res}). A range of subhalo masses are used to examine how the mass of both the primary and secondary impacts tidal stripping.

The results of these calculations are in Fig.~\ref{figtrunc}, which depicts the dependence of the satellite truncation radius (in units of the scale radius) on the radial distance from the centre of its host halo. Decreasing values of $\rtrs$ indicate more tidal stripping because, as the tidal radius approaches the scale radius of the satellite, an increasing amount of mass is stripped. The maximum circular velocity occurs at a scale radius $\rs = 2.2$; so when $\rtrs \sim 2$, internal structure is affected. Satellites of varying masses are shown with different colours. Smaller haloes collapse earlier, increasing their concentration. This results in a larger density at $\rs$, causing the less massive satellites to be more resilient against tidal stripping.

Fig.~\ref{figtrunc} includes results from an NFW main halo ($\Mvir=1.4\times10^{12}~\Msun$) with and without a fraction of the baryons ($M_{\rm disc}=6\times 10^{10}~\Msun$) coalesced at the centre. The pure NFW halo mimics cosmological simulations, where baryons are assumed to trace the DM in an unbiased way. The impact of including a disc component is clearly seen for orbits with pericentres within $\sim 30$~kpc. At this radius, each satellite's tidal radius decreases by $\sim14$~per cent, which is a significant effect. For orbits with pericentres within $\sim 20$~kpc, the satellites will be destroyed under the influence of a MW with or without a disc component because $\rtrs \la 1$.  In contrast, if the pericentre is greater than 100~kpc, then $\rtrs > 4$ regardless of whether a disc component is included or not, which produces weak tidal stripping.

For the regime of orbits with $30~\kpc \la r \la 100~\kpc$, the inclusion of a disc has a more subtle but important effect. From $30~\kpc \la r \la 60~\kpc$ the satellites have obviously strong tidal stripping if the MW disc is included. Orbits with $60~\kpc \la r \la 100~\kpc$ may also undergo increased tidal stripping and should be examined further. We predict that the inclusion of a disc for orbits with these pericentres results in an important regime of mass-loss and circular velocity decrease. As such, we have created {\it N}-body simulations of satellites on elliptical orbits with pericentres between 30 and 70~kpc, to investigate this region.

\subsection{Effects of tidal stripping in {\it N}-body simulations}
\label{subsec:tidalstripping}
Each of our three dwarf galaxy mass removal cases had 20~per cent of its mass removed, but they produced different maximum circular velocity values and central density profiles. Instantaneous mass removal (IR) and exponential mass removal (ER) scenarios both resulted in a 20~per cent reduction in $\vmax$ due to adiabatic expansion. The pre-halo formation mass removal (PR) scenario did not undergo adiabatic expansion; therefore, its $\vmax$ only decreased by 10~per cent in isolation. See Section~\ref{subsec:baryonremoval} for the mass removal description and results.

After each dwarf evolved in isolation, they were put into a variety of circular and elliptical orbits around a MW-like host as described in Section~\ref{subsubsec:setuptidalstripping} and listed in Table~\ref{tab:2}. Comparing the orbital evolution of these three haloes will provide insight into how pericentre, number of orbits completed, and central density of the satellite impact their structural and $\vmax$ evolution.

\begin{figure}
\includegraphics[trim= .8cm 2cm 1.5cm 1.7cm,clip,width=0.37\textwidth,angle=90]{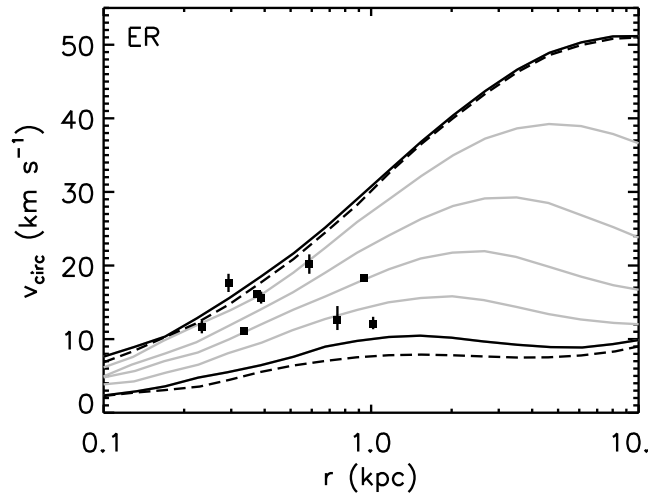}
\caption{Evolution of the circular velocity profiles for two mass removal satellites throughout their 50~kpc circular orbits. The exponential mass removal (ER) satellite is shown in solid lines. The upper and lower black lines indicate its initial and final states, respectively, while the grey lines in between are each separated by 1~Gyr. After 5~Gyr, the satellite's $\vmax$ has been stripped from 50 to $10~\kms$. The initial and final state of the instantaneous mass removal (IR) case is shown as dashed lines for comparison. The black squares are observational data points of the half-light velocities for the classical LG dSphs from \citet{Walker:2009ez}.}
\label{fig:vcircevolution}
\end{figure}

\begin{figure*}
\includegraphics[trim= .8cm 2cm 1.5cm 1.7cm,clip, width=0.4\textwidth,angle=90]{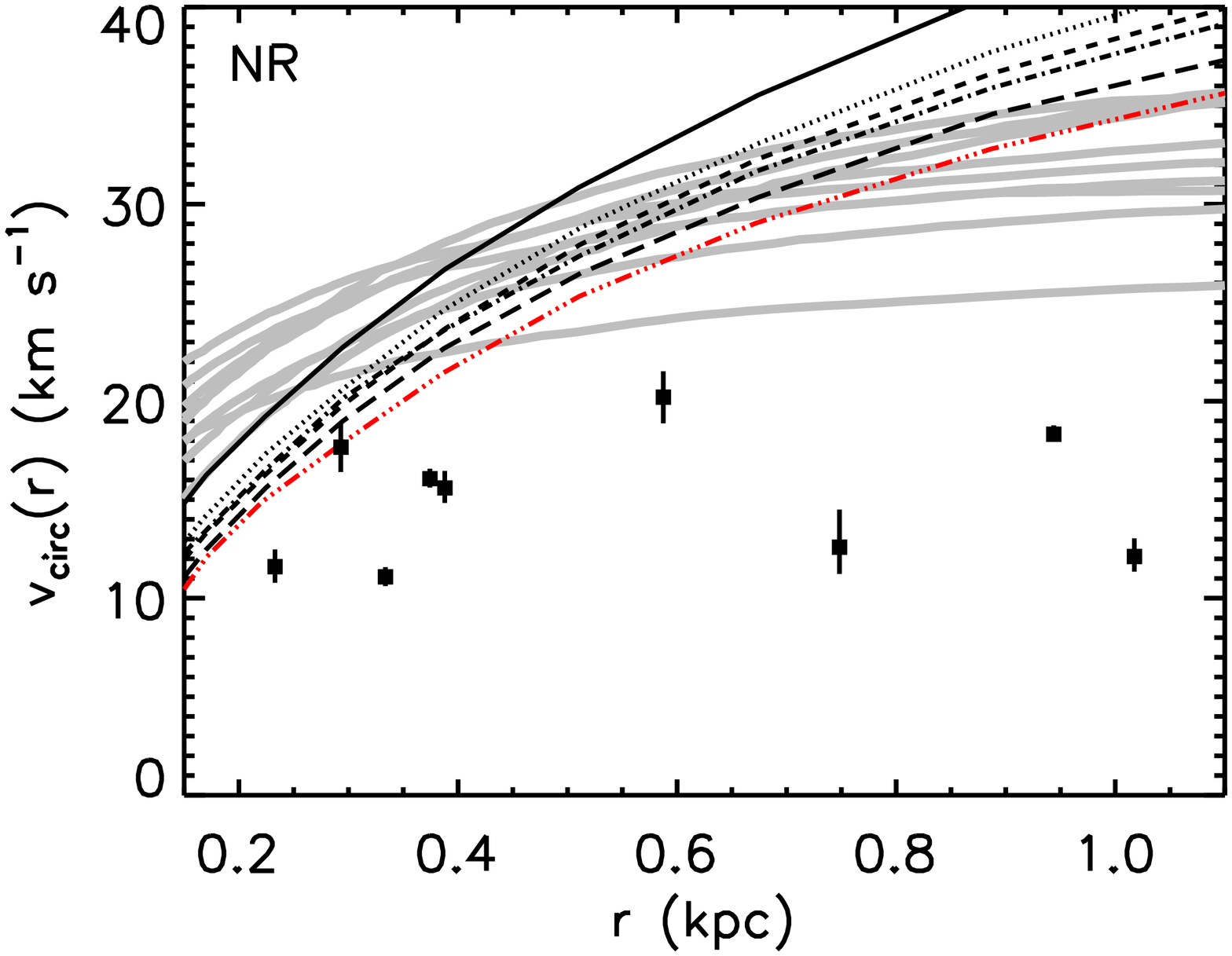}
\includegraphics[trim= .8cm 2cm 1.5cm 4.6cm,clip, width=0.4\textwidth,angle=90]{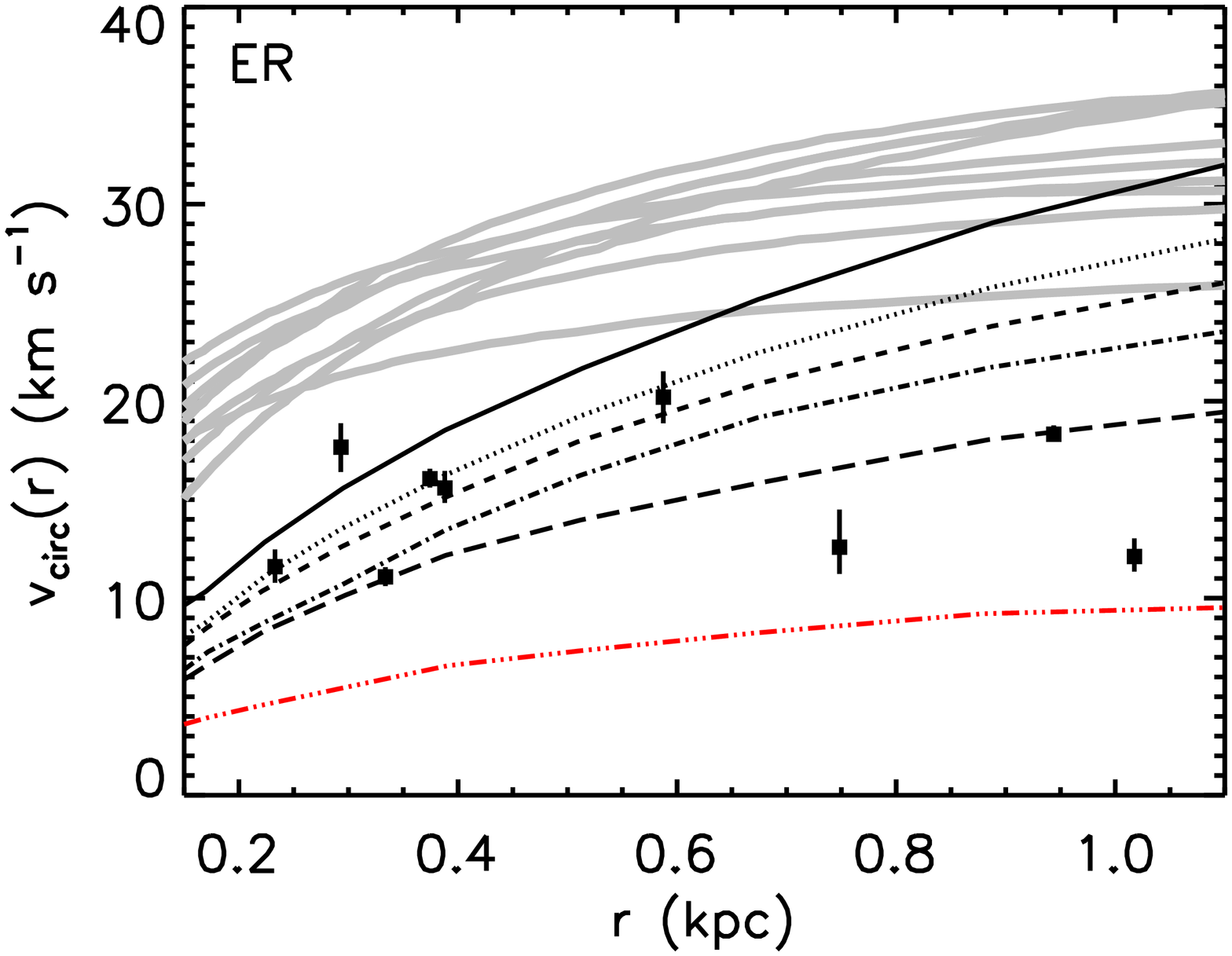}
\caption{Comparison of our elliptical orbits to observations of the MW satellites from \citet{Walker:2009ez} (black squares) and the Aquarius E halo's `massive failures' from the analysis of \citet{BoylanKolchin:2012id} (thick grey lines). Left: the circular velocity profiles of the NR case. This profile does not agree with the MW dSph population and was created to mimic the Aquarius cosmological simulations by not including the effects of baryons. Right: the circular velocity profiles of the exponential mass removal (ER) case.  These profiles span a much larger range of parameter space and are in agreement with observed MW dwarfs due to their inclusion of baryonic effects. Both panels show the initial isolated profile as a solid black line. The elliptical orbits after 5~Gyr of evolution are shown from top to bottom as 150--70~kpc (dotted), 150--50~kpc (short dashed), 150--30~kpc (dot--dashed) and 120--30~kpc (long dashed). The 50~kpc circular orbit is shown as the bottom most triple dot--dashed line (red) for reference of the most dramatic orbital evolution of our simulations.}
\label{fig:bkcomp}
\end{figure*}

\subsubsection{Circular orbits}
\label{subsec:circorbits}
Four different dwarfs were placed in circular orbits around the MW perturber. The NR dwarf, which had evolved in isolation, was placed around a pure NFW perturber to mimic the results of DM-only simulations. The three mass removal dwarfs -- two that showed adiabatic expansion (IR and ER) and one without (PR) -- were placed in circular orbits around a MW perturber with an NFW and spheroidal exponential `disc' component. All satellites ran in their orbits for 5~Gyr with initial and final maximum circular velocities listed in Table~\ref{tab:2}.

We investigated two orbital regimes (i) at $r \ge 100\, \kpc$ where the mass-loss was expected to be low based on our truncation radius analysis ($\rtrs > 4$) and (ii) at $30\, \kpc < r < 100\, \kpc$ where the truncation radius ratio had values between $2 < \rtrs < 5$. This second regime is a sensitive area where a large amount of mass may be lost. Pericentre passages within 30~kpc, $\rtrs < 2$, would result in extreme tidal stripping for the satellite regardless of its mass or central density.

To investigate the first regime, circular orbits at 150 and 100~kpc were created. The truncation radius to scale radius ratio ($\rtrs$) is large at this orbital distance (NR with a pure NFW MW: $\rtrs =8.7$ at 150~kpc and $\rtrs=6.3$ at 100~kpc). Due to the large $\rtrs$ value, very little matter would be stripped from the satellite even after a long time period in these orbits. This was confirmed by our results listed in Table~\ref{tab:2}, where the 150~kpc orbits for the NR case showed only a 5~per cent change in the satellite's $\vmax$ after 5~Gyr of orbital evolution.

Even in distant orbits, the combined effects of including baryons in simulations, namely mass removal with subsequent adiabatic expansion and the inclusion of a baryonic disc in the host, doubled the reduction in $\vmax$ from 5 to 10~per cent (for 150~kpc orbit) and 11 to 22~per cent (for 100~kpc orbit), as given in Table~\ref{tab:2}. Mass-loss was even more significant than the change in $\vmax$, with losses of 60~per cent in a 150~kpc orbit and 80~per cent in a 100~kpc orbit for the ER case. This shows that the evolutionary history of galaxies in pure DM simulations may produce incorrect subhalo masses at redshift zero, even if they do not interact closely with their hosts.

Closer circular orbits at 70 and 50~kpc were run to examine more extreme tidal stripping. Although the haloes had different $\rtrs$ values, all underwent significant mass-loss during 5~Gyr. The most dramatic of these were the ER and IR 50~kpc orbits, where $\rtrs = 2.2$. Fig.~\ref{fig:vcircevolution} displays the significant tidal stripping satellites experience when orbiting close to their host, by examining their circular velocity profiles. The ER case is shown every Gyr throughout its 50~kpc circular orbit, and the initial and final IR profiles are shown for comparison. This allows one to trace the evolution of the entire circular velocity profile for this specific case. The $\vmax$ and $\rmax$ of the ER case decreased with time until the final satellite has $\vmax = 10~\kms$, with a significant fraction of bound mass existing as tidal features. At 2~Gyr, the $\vmax$ has decreased by 40~per cent, converting a 51~$\kms$ subhalo to a much smaller 30~$\kms$ satellite. Although the ER and IR cases began with similar circular velocity and density profiles, the final IR case is slightly more disrupted than the ER subhalo, which retained a small satellite core.

Observational data points of the half-light velocities for the classical LG dSphs are shown in Fig.~\ref{fig:vcircevolution} as well. The initial ER profile is too massive to be consistent with the majority of LG dSphs but is consistent at  $2\sigma$ with Draco, the data point just above it, and Leo II, the data point with the smallest radius value. At the end of 5~Gyr of evolution, the satellite is capable of matching all observed LG dSphs at some point in its evolutionary history. The sudden expulsion of baryons combined with orbiting a MW potential with a disc component has converted a massive satellite, incapable of matching any of the LG dSphs, into a dwarf that matches a given observed dSph at a fixed time in its evolution. As seen in the evolution of the circular velocity profiles, satellites with $\rtrs = 2.2$ are capable of undergoing extreme disruption and possibly becoming completely destroyed. We will further examine the tidal disruption for different satellites in our elliptical orbits.

\begin{figure*}
\centering
\includegraphics[trim= 3.5cm 2cm 1.5cm .8cm,clip, width=0.32\textwidth,angle=90]{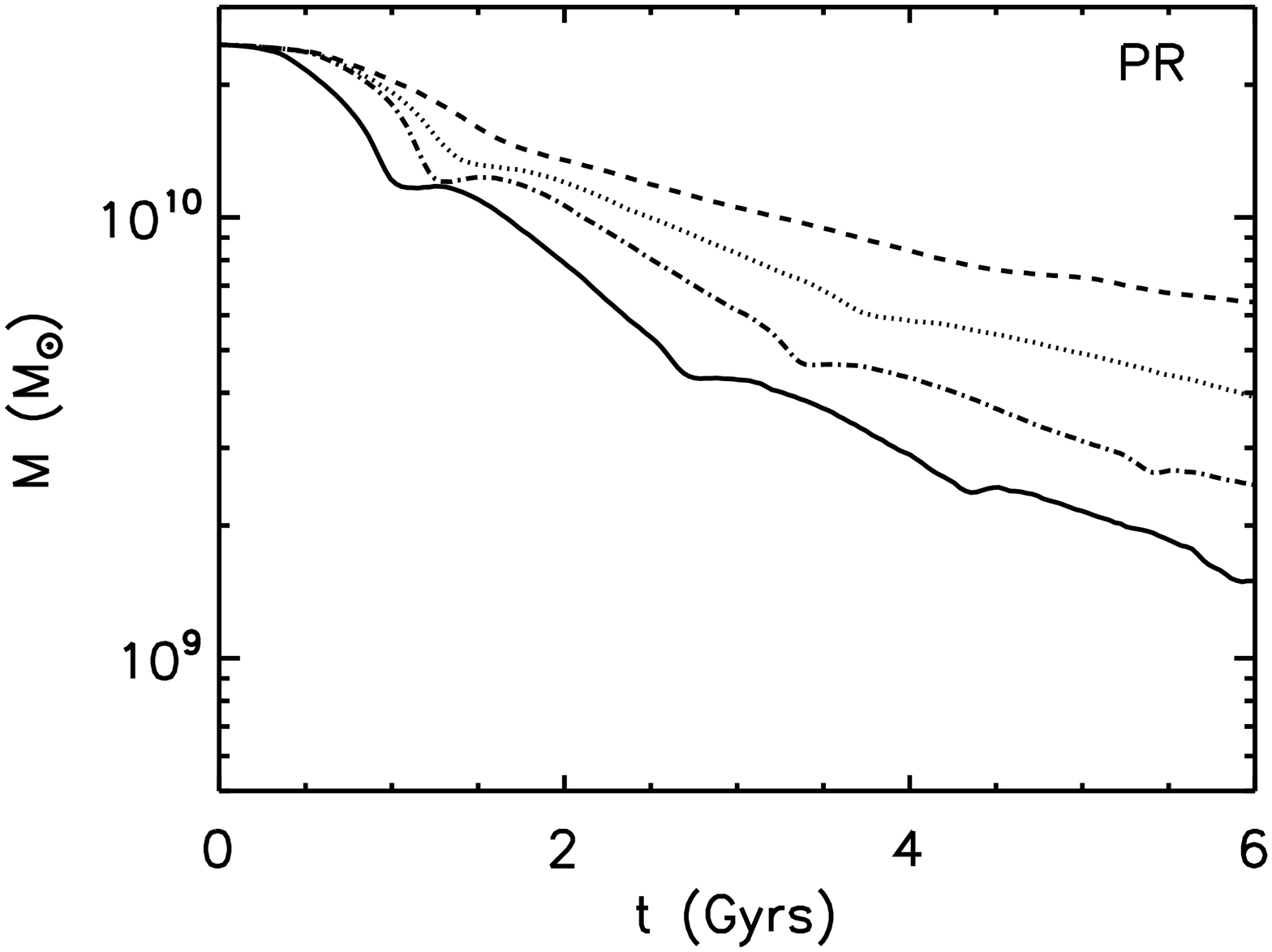}
\includegraphics[trim= 3.5cm 2cm 1.5cm 4.6cm,clip, width=0.32\textwidth,angle=90]{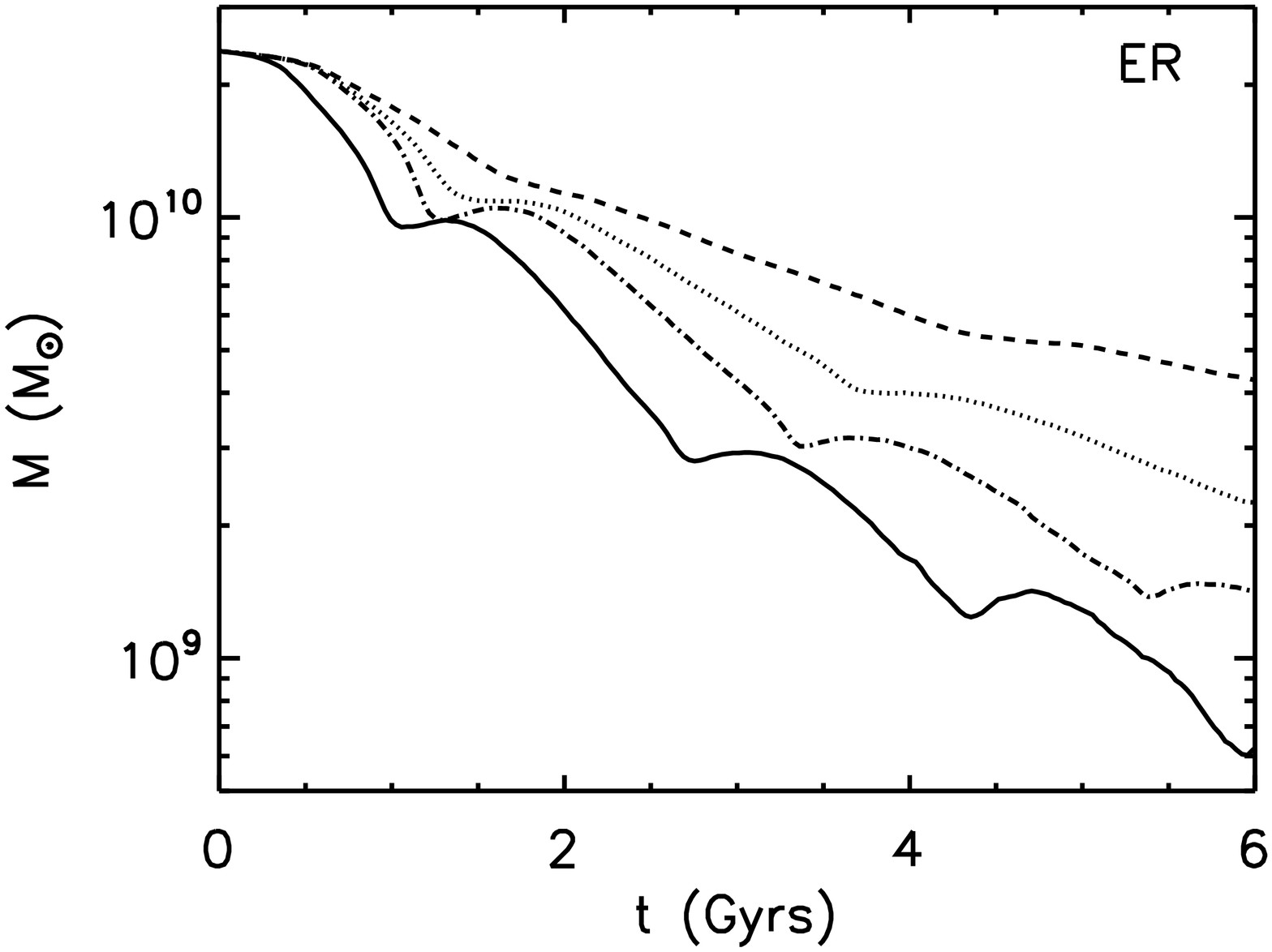}
\includegraphics[trim= .8cm 2cm 1.5cm .8cm,clip, width=0.382\textwidth,angle=90]{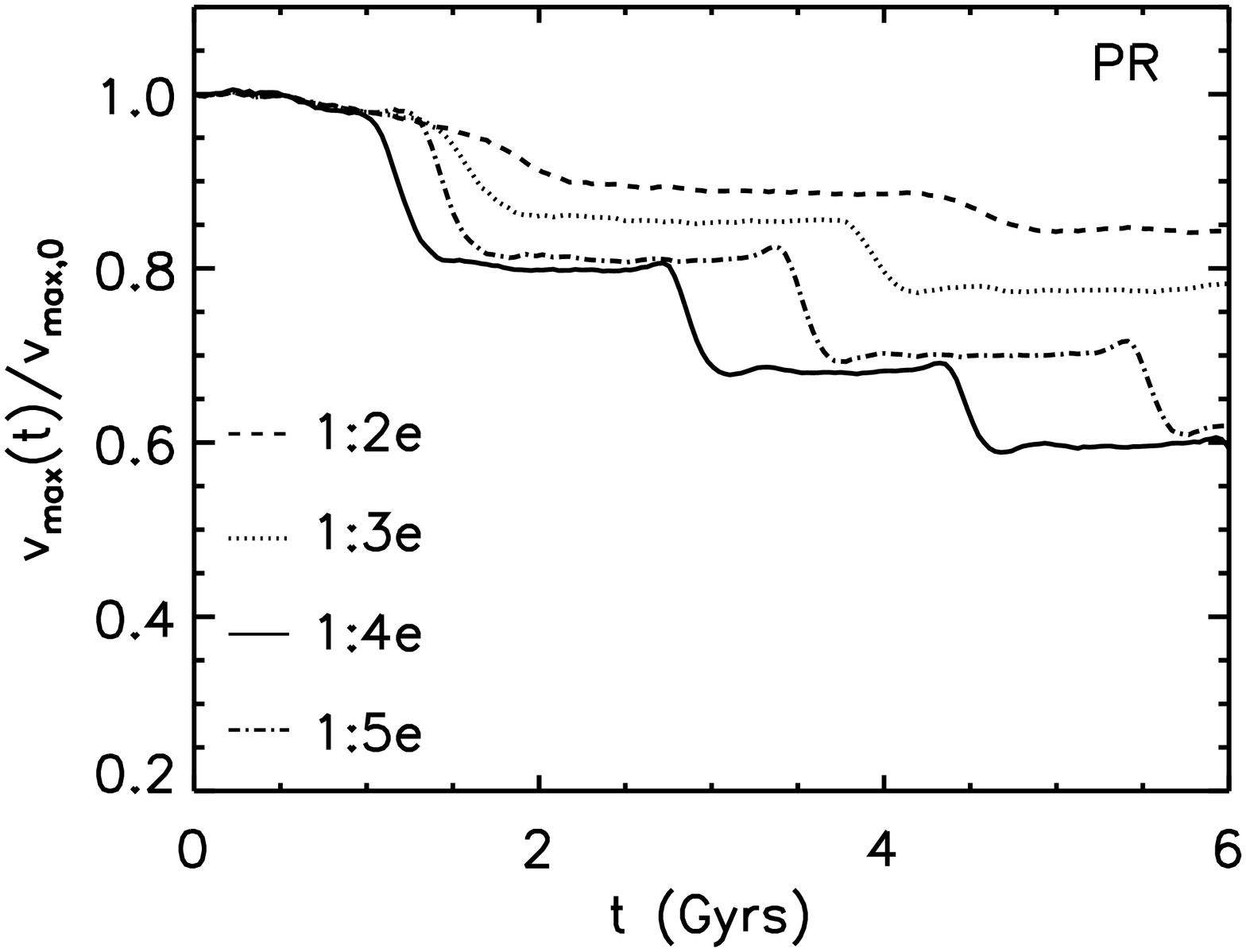}
\includegraphics[trim= .8cm 2cm 1.5cm 4.6cm,clip, width=0.382\textwidth,angle=90]{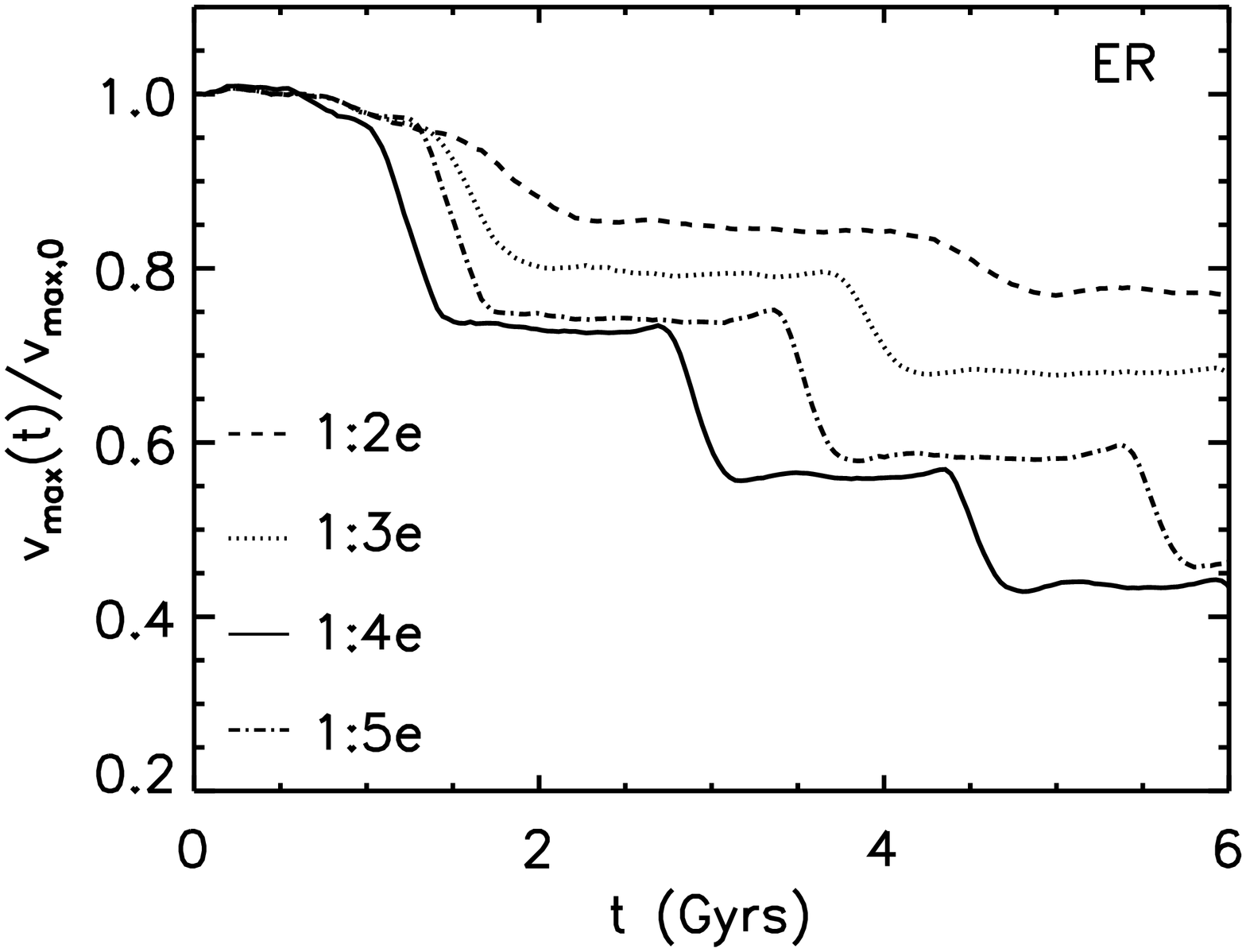}
\caption{Mass evolution and fractional change in maximum circular velocity ($\vmax/v_{\rm max, i}$) for the pre-halo formation mass removal (PR) satellite, on the left, and exponential mass removal (ER) satellite, on the right. The four elliptical orbits: 1:2e (150~kpc to 70~kpc), 1:3e (150~kpc to 50~kpc), 1:4e (120~kpc to 30~kpc) and 1:5e (150~kpc to 30~kpc) are shown. Mass continually decreases throughout the 6~Gyr of evolution. Pericentre passages occur at the steep drops in both mass and $\vmax$. The 1:4e and 1:5e orbits have the same pericentre distance of 30~kpc, with the difference being their apocentres. The 1:5e orbit experiences less mass and $\vmax$ loss at a given time because of the longer time to complete one orbit. If this time-scale is taken into account, the same pericentre distances create very similar evolutionary histories.}
\label{fig:difforbits}
\end{figure*}

\begin{figure}
\includegraphics[trim= 3.4cm 1.5cm 1.5cm 1.7cm,clip,width=0.31\textwidth,angle=90]{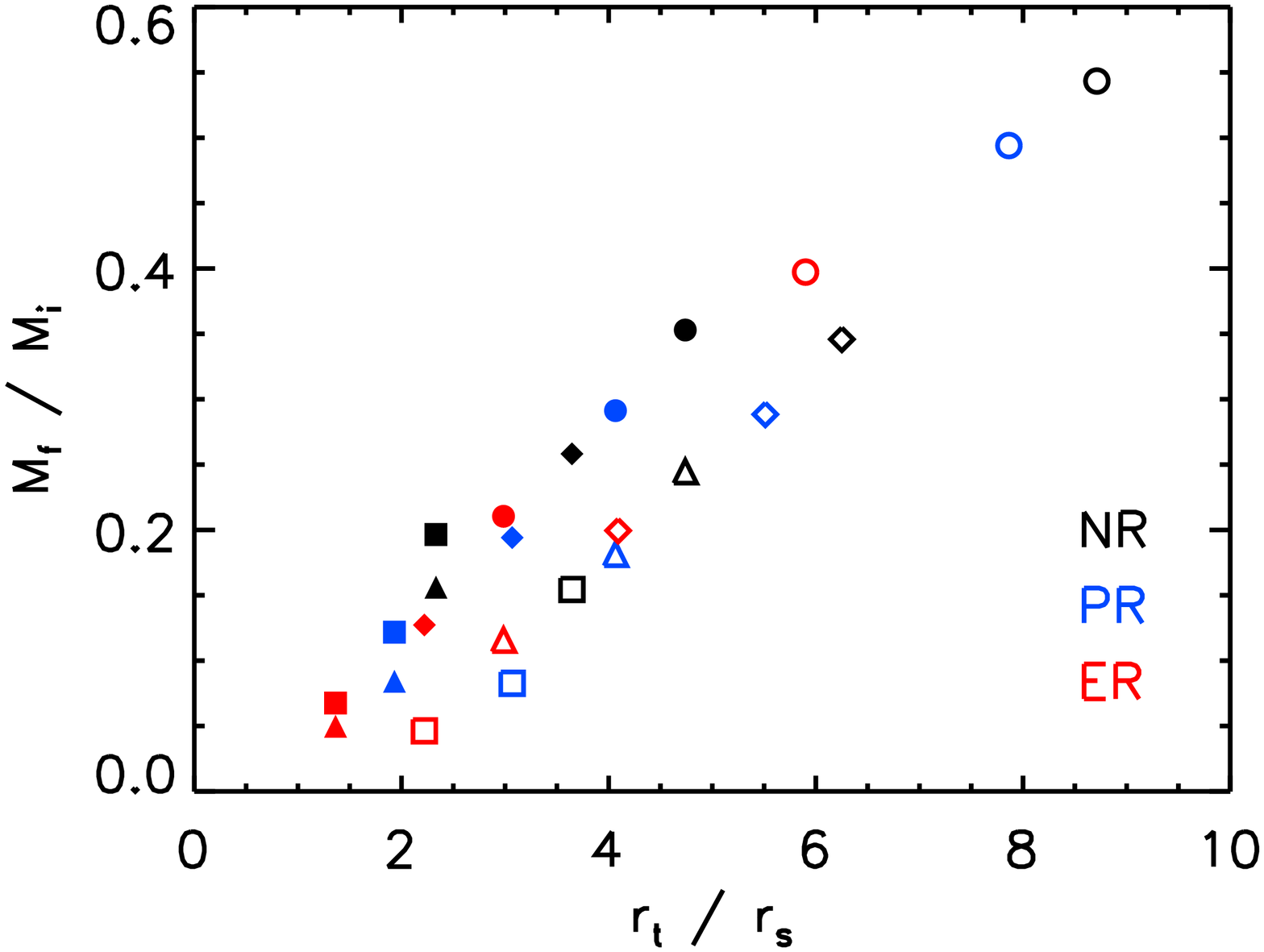}
\includegraphics[trim= .8cm 1.5cm 1.5cm 1.7cm,clip,width=0.367\textwidth,angle=90]{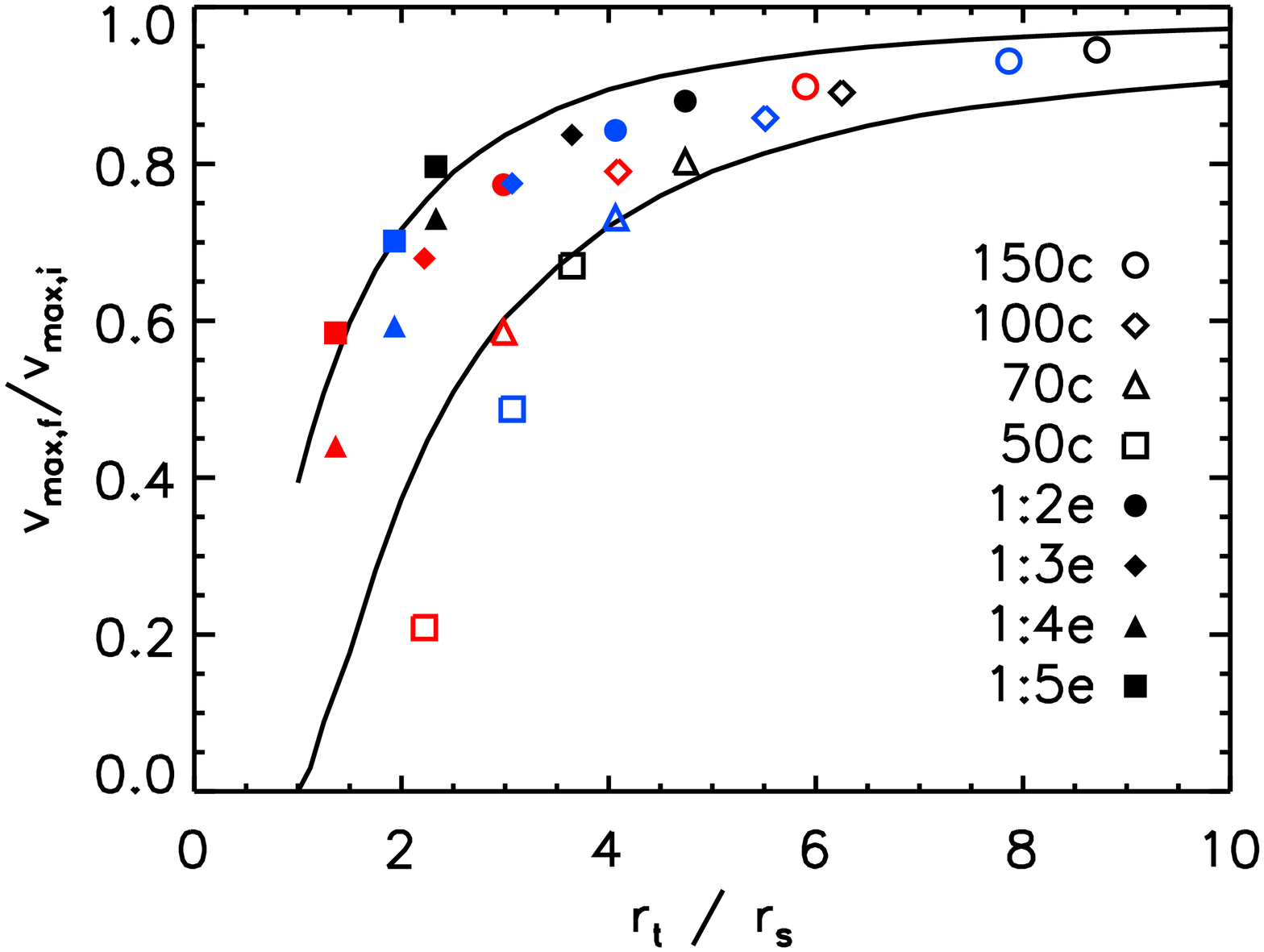}
\caption{Results of tidal stripping for the mass and $\vmax$ of a satellite after 5~Gyr. Symbols indicate different orbits with open symbols for circular orbits and filled symbols for elliptical orbits. The colours indicate different satellite cases: red for exponential removal, blue for pre-halo formation removal and black for no removal. Different satellite cases (colours) with the same orbit (symbol) always occur from left to right as ER, PR and then NR, with increasing values of $\rtrs$. The fraction of mass retained increases linearly with increasing $\rtrs$ so that the NR case always has the largest final mass fraction. When $\rtrs$ is below 4, tidal stripping is very important, as seen by the significant mass-loss and decrease in $v_{\rm max, f}/v_{\rm max, i}$. The solid black lines display our model for tidal stripping by iteratively removing all unbound particles and recalculating the tidal radius. The results of 10 and 30 iterations are the top and bottom solid lines, respectively.}
\label{fig:vmax_rt}
\end{figure}

\subsubsection{Elliptical orbits}
\label{subsec:elliptorbits}
All four mass removal cases were placed on elliptical orbits and ran for 6~Gyr. These orbits have apocentre--pericentres of 150--70, 150--50, 150--30, and 120--30~kpc. As discussed in Section~\ref{subsubsec:setuptidalstripping}, these orbits were selected in order to examine an interesting range of truncation radii ($1 < \rtrs < 5$) and be realistic generalizations of classical LG dSphs orbits. The initial and final maximum circular velocities for the elliptical orbits are listed in Table~\ref{tab:2}.

As a comparison between the satellites run in elliptical orbits, Fig.~\ref{fig:bkcomp} shows the circular velocity profiles of the NR dwarf (left-hand panel) and the ER dwarf (right-hand panel). We plot our results for the isolated NR and ER haloes as well as the results of all elliptical orbits and the 50~kpc circular orbit at 5~Gyr. This figure also includes observational data points of the half-light velocities for the classical LG dSphs and lines from \citet{BoylanKolchin:2012id} fig.~3 for the Aquarius E halo.

The initial NR subhalo was created in order to match cosmological simulations, without baryon removal and without a MW disc component. The circular velocity profile for the NR halo does not agree within $2\sigma$ with observed MW dSphs, mimicking the high mass of the Aquarius `too big to fail' satellites. The initial ER satellite is in better agreement with observations because it includes baryonic mass removal before orbital evolution. This demonstrates the importance of baryonic removal in creating agreement between cosmological simulations and observations. The magnitude of the initial halo's transformation depends on the amount and timing of mass removal from the halo.

In agreement with the results of \citet{BoylanKolchin:2012id}, many elliptical and circular orbits for our `cosmological' simulation of the NR halo remained inconsistent with observations of MW dSphs after 5~Gyr of orbital evolution. Only with an extremely close circular orbit (50~kpc) or a 120--30~kpc elliptical orbit does the NR satellites agree with Draco at $1\sigma$.

The results for the ER halo are in much better agreement with the entire population of LG dSphs. As the ER satellite included both mass removal and a MW disc component, we find it is critical to compare observed dSphs to simulations including baryonic effects. This is true regardless of the orbit chosen, as even distant elliptical orbits around a MW with a disc would undergo substantial tidal stripping.

The PR halo was formed with 80~per~cent mass of the NR halo in equilibrium so it did not undergo adiabatic expansion. Due to its larger initial circular velocity profile than the ER case (as seen in Fig.~\ref{fig:equilib}) it undergoes less dramatic tidal evolution. After 5~Gyr, this halo agrees with Draco on its 150--70, 150--50 and 150--30~kpc elliptical orbits. Additionally, it is in agreement with the initial circular velocity profile of the ER halo on its 120--30~kpc elliptical orbit after 5~Gyr and the 150--30~kpc orbit after 6~Gyr.

The importance of including the MW disc is seen from these results as tidal stripping significantly reduces the mass and $\vmax$ of the dwarfs. By combining baryon removal and orbital evolution around a MW including a disc component our initially inconsistent dwarfs are able to match the observed LG dSph population. However, we find that simply including a disc component in the MW is not sufficient to create agreement between our dwarf and the LG dSph population. The PR and ER haloes have the same total mass and evolve on the same orbits, meaning their internal structure drives the differences between their final profiles. Baryonic removal from already formed haloes must be included in addition to orbital evolution around a MW with a disc component in order to match the entire LG dSph population.

Fig.~\ref{fig:difforbits} shows the evolution of mass and fractional maximum circular velocity ($\vmax/v_{\rm max, i}$) for all PR and ER elliptical orbits over 6~Gyr. As the satellites began their orbits, they immediately lost a substantial amount of mass under the tidal influence of the MW. Most dwarfs lost half their mass at their first pericentric passage and by their second pericentre some satellites had only one fifth of their mass remaining. 

Pericentres are marked by `steps' or `bounces' in the circular velocity and mass evolution, respectively. Although, mass is lost continuously throughout an orbit, $\vmax$ only changes significantly at each pericentre. The slight increase in mass at pericentre is due to mass in the tidal features temporarily becoming bound to the satellite. We found $\vmax$ to be a more robust parameter for describing the transformation a satellite has undergone throughout its orbit.

As expected, orbits with closer pericentres experience more dramatic tidal stripping as seen from their mass-loss and reduction in $\vmax$. Interestingly, the 150--30~kpc (1:5e) and 120--30~kpc (1:4e) orbits have approximately the same change in mass and $\vmax$ after the same number of pericentre approaches. Although they have different apocentres and ellipticities, the similarity between their tidal evolutions shows that pericentre distance, and consequently tidal radius, is a good indicator of total tidal stripping.

A more accessible parameter for observers is a measurement of the circular velocity within the central region of the galaxy. The evolution of circular velocity at 500 pc proceeds similarly to the $\vmax$ evolution. The circular velocity steps down with each pericentric passage and remains constant for the remainder of the orbit. The initial $v_{500}$ has a value of 21.5~$\kms$ for the ER case and has final values in the range of 14 to 19~$\kms$ for 6~Gyr of evolution in elliptical orbits.

Throughout their orbital evolution, the satellites lose most of their mass from the outer regions, steepening the outer density profile. The inner regions are less susceptible to tides and retain their cuspy shape even after significant mass-loss, in agreement with \citet{Hayashi:2003kz} and \citet{Penarrubia:2010du}. Additionally, the density profile of each satellite develops a break radius feature in the outer regions, which evolves with time \citep[see][]{Penarrubia:2008eq,Penarrubia:2009kp,Penarrubia:2010du}.

The structure of the halo is determined by the amount of mass lost \citep[as discussed in][]{Hayashi:2003kz,Penarrubia:2008eq,Penarrubia:2010du}. We have found that pericentric distance and tidal radius are able to predict the tidal stripping of a halo. In order to examine this in detail, we investigated how changing the ellipticity and pericentre affected mass removal, based on $\rtrs$. In Fig.~\ref{fig:vmax_rt}, we show the fractional change in mass and $\vmax$ after 5~Gyr in relation to the $\rtrs$ of each satellite. We calculated the truncation radius of our subhaloes at pericentre in the same method as used in Section~\ref{analytics}. Circular orbits have radii of 150, 100, 70 and 50~kpc, and elliptical orbits have pericentres of 70, 50 and 30~kpc.

We input the analytic form of the MW mass, $M(R)$, the pericentre, $R$, and the numerical isolated satellites' mass profile, $m(r)$, into equations~(\ref{eq:truncrad}) and (\ref{eq:truncrad_res}). The minimum of the tidal and resonant radii for each satellite and pericentre combination was taken as the truncation radius, $\rt$ and divided by the satellites' $\rs$ to create the value $\rtrs$.
 
In Fig.~\ref{fig:vmax_rt}, we examined how the tidal radius impacts the mass-loss after 5~Gyr of evolution.  The satellite cases are plotted in different colours and symbols are used for different orbits, as listed within the figure. Note that the instantaneous removal case is not included as its tidal stripping and $\rtrs$ values are almost identical to the exponential removal case.

It is apparent that for satellites with large pericentres, $\rtrs > 4$, the tidal stripping is minimal. As expected, over half of a satellite's mass is lost, but the $\vmax$ does not suffer a similar reduction and remains above 0.8 the initial value for these cases. For example, the shift in $\rtrs$ between the NR~150c and ER~150c points is due to the adiabatic expansion of the mass removal haloes, which increases their $\rs$ values. This gives smaller $\rtrs$ values, while only slightly lowering their $\vmax$ fraction.

However, for satellites on orbits that reach $\rtrs \sim 4$ and below, there is a broad range of possible mass-loss scenarios. In this regime, a satellite's mass can range from 0.3 to 0.05 times the original total mass, and the circular velocity can drop from 0.8 to 0.2 its original value. This is a dramatic decrease, indicating that tidal radius over scale radius at pericentre is an important parameter for predicting mass-loss.

The $\rtrs$ values of the PR haloes are significantly larger than those of the ER haloes, as denoted by different colours with the same symbols. The internal structure of the halo makes the PR cases resistant against tidal stripping so that the satellite does not experience as large a reduction in $\vmax$ or mass.

As circular orbits are always at the same distance, they are able to reduce their mass and $\vmax$ more during a fixed time than an elliptical orbit with the same pericentre. Points with the same $\rtrs$ value indicate that they have the same pericentre making for an easy comparison between points that lie on vertical lines. For example, the 70~kpc circular orbit (70c) and the 150--70~kpc elliptical orbit (1:2e) have the same $\rtrs$ for each satellite case of a single colour. The vertical separation between these pairs of points is due the length of time that the satellites are near the primary.

Although satellites on circular orbits undergo a larger number of orbits around the MW in the same amount of time as the elliptical orbits, the tidal forcing they feel is constant throughout all time. This means that tidal truncation is the main method of mass removal, without tidal shocks. We have included our tidal stripping model, as described in Section~\ref{subsec:atidalstrippingmodel}, that mimics tidal truncation. Unbound mass is iteratively removed from the satellite, and the tidal radius is recalculated with each iteration. The results of this model with 10 and 30 iterations are plotted in Fig.~\ref{fig:vmax_rt}. Our simulation results typically fall between these two lines, indicating that the number of truncation episodes, a proxy for the number of pericentric passages, can account for the spread between points with the same $\rtrs$ value.

\section{Discussion}
\label{sec:discussion}

In this paper, we focus on examining if the inclusion of baryonic effects is able to reduce the discrepancy between simulations of DM subhaloes and observations of dSphs.

The progenitor presented in this work is likely to have a $M_* \sim 10^{7}~\Msun$ for the initial $10^{10}~\Msun$ virial mass. We show that this halo undergoes a severe reduction in $\vmax$ and $\Mvir$ depending on its orbit and method of baryon removal. Our simulations can produce Fornax like objects with significant $\Mvir$ reduction during tidal stripping and little evolution in $M_*$.

Less luminous dSphs must have progenitors with far lower total stellar mass than the cases shown here. This may occur when dIrrs are subjected to gas removal at an early epoch that shuts off SF and results in low stellar masses today. As an example, we examine Draco and Ursa Minor, which have $M_* \sim 3\times 10^5~\Msun$ \citep{McConnachie:2012fh}, are at similar distances $D_{\rm MW} \sim 76~\kpc$ \citep{McConnachie:2012fh} and are expected to inhabit haloes with $\vmax \sim 25~\kms$. Both show SF occurring at early times and shutting down 10~Gyr ago \citep{Grebel:1998uz}, implying their accretion around redshift two. Therefore, the low stellar mass of objects such as Draco is a result of the shut-down of SF after early accretion on to the MW.

In our simulations, we show how a gas-rich dwarf may undergo significant evolution by including the effects of baryons. We are not advocating that only objects with $\Mvir = 3\times10^{10}~\Msun$ are able to create dSphs, but that baryonic effects can significantly alter the evolutionary history of satellites and must be taken into account. We illustrate the power of baryon removal and tidal stripping when including a disc in the MW and caution against using collisionless simulations when claiming discrepancies between observations and simulations of small-scale structure. Our conclusion is therefore that baryons must be included when examining the `too big to fail' problem.

Our baryon removal mechanism in the ER and IR cases relies on gas removal from a formed galaxy. Alternatively, the PR halo represents the case where a halo formed in a baryon deficient environment. After 5~Gyr of evolution in a 150 to 30~kpc elliptical orbit the NR, PR and ER haloes had $\vmax = 51$, $40$ and $30~\kms$, respectively. Baryon removal and the inclusion of a MW disc component are important parts to resolving the presence of substructures in simulations that are too massive to host LG dSphs.

Sudden baryon removal is more efficient in transforming a massive dwarf galaxy into a low-mass satellite than baryon removal before the halo forms. A likely culprit for sudden removal is ram pressure stripping of a dIrr as it falls into a host halo.

Due to the environmental requirement for the dIrr--dSph transformation, isolated dIrrs may hold the key to solving the `too big to fail' problem. If so, a field dIrr would have the same stellar mass within the tidal radius as a dSph; however, it would also include a gaseous component and a larger DM density. Ram pressure stripping would remove the gaseous component, causing the halo to adiabatically expand, and DM and stars would subsequently be stripped throughout its orbital evolution, a theory we examined in our hydrodynamical and IR and ER {\it N}-body simulations.

Observational evidence for different dSph progenitor models may be found by examining the central densities of dIrrs. Gas-rich field dwarfs should have higher central DM densities than dSphs as they have not undergone tidal stripping or baryon removal. However, the physical properties of the dSphs progenitors are unknown. Furthermore, there is a large intrinsic spread in concentrations and central DM densities for a given halo mass. This makes it extremely difficult to compare dIrrs and dSphs.

This has recently been examined in a statistical sense by \citet{Ferrero:2012bt}. Using published $\HI$ data for a range of stellar masses, they compare the enclosed mass of dwarf galaxies and find that they inhabit lower mass haloes than expected from abundance matching and semi-analytic models. These results can be understood if the $\HI$ measurements underestimate the circular velocity of the galaxies or if the abundance of dwarfs with stellar masses below $10^{8.5}~\Msun$ has been strongly underestimated. Further examination of isolated dIrrs is important to confirm or deny the baryon removal models implemented in this paper.

An additional complexity in the study of substructure is the central density of satellites. We have not examined core creation in these simulations, as this requires repeated mass blowout events (see Section~1 for more discussion). The final density profiles of our satellites were steepened in the outer regions where mass was preferentially removed. In the inner regions, the density profile remained cuspy and retained the initial NFW shape. The creation of cored profiles has been found to resolve the `too big to fail' problem \citep{Brooks:2012wa,Zolotov:2012hi} and would agree with recent observations of LG dSphs \citep[e.g.][]{Walker:2011eg,Amorisco:2012fp}.

\section{Conclusions}
\label{sec:conclusions}
One of the goals of this paper is to resolve some of the small-scale problems of the standard $\Lambda$ cold dark matter model, namely the apparent overabundance of satellites with $30~\kms < \vmax < 50~\kms$ \citep{BoylanKolchin:2011ky, BoylanKolchin:2012id}. In line with other recent results of cosmological simulations of galaxy formation \citep{Brooks:2012wa, Zolotov:2012hi}, we argue that the solution of the problem is in the gas physics of forming dwarf galaxies. By posing the question `what are the effects of baryons on dwarf galaxies,' we hope to examine just one aspect of the increasingly complicated fields of galaxy formation and evolution.

Most of the existing theoretical predictions and constraints on the abundance and structure of small satellites (substructures) are made using {\it N}-body simulations. Because observed dwarf galaxies in the MW are heavily dominated by the DM, one expects that there is little impact of baryonic physics on the structure of dwarf satellites. However, this appears not to be true; the corrections due to baryons are significantly larger than one would naively expect.

Baryon removal may reduce the maximum circular velocities of satellites enough to explain the `too big to fail' problem. The central mass distribution of a dwarf may be substantially altered both by gas removal at infall and by tidal stripping due to the host galaxy. {\it N}-body simulations underestimate stripping because (i) baryons are not removed from the dwarf and (ii) the mass of the host is not enhanced by infalling baryons. Accounting for these effects can alter the circular velocity profiles of the subhaloes, such that they are `small enough to succeed' in explaining the observed structure of the MW dSphs.

We used a number of methods to investigate how baryonic physics impacts the structure of a dwarf spheroidal. We ran hydrodynamical cosmological simulations, high-resolution {\it N}-body simulations and an adiabatic expansion model to examine the effects of baryon removal from dwarf galaxies. The impact of tidal stripping on dwarf galaxies was explored using a series of high-resolution {\it N}-body simulations and two analytical models.

Our two high-resolution cosmological simulations were run with and without gas removal. The gas removal case was created to reproduce the heating and removal of gas as a dIrr falls into a host galaxy. This dwarf expanded adiabatically in response to the mass-loss. Baryon removal from the {\it N}-body simulations also resulted in adiabatic expansion for our exponential and instantaneous mass removal cases. The rate of removal of the gas does not affect the final results. In other words, both fast and slow removal provide the same amount of expansion.

The gas removal results were modelled with the \citet{Blumenthal:1986ie} prescription for adiabatic expansion. Mass removal causes dwarfs to both expand and lower their circular velocities. The magnitude of the expansion depends on the fraction of mass removed and on the slope of the density. In the case when the same fraction of mass is removed independent of radius, the shape of the density profile does not change. The maximum decline in the density, according to adiabatic expansion model, is equal to $(1-\fb)^4$, where $\fb$ is the cosmological fraction of baryons. This decline is expected to happen for galaxies with constant density cores. For galaxies with cusps, the decline is smaller. For the NFW $\beta = 1$ cusp, the decline is still substantial with $\rho_\rmf/\rho_\rmi = (1-\fb)^3$. If we assume that the mass removed is the cosmological average fraction, then $\rho_\rmf/\rho_\rmi \approx 0.5$.

We tested a series of circular and elliptical orbits for each dwarf with high-resolution {\it N}-body simulations. Without baryon removal, tidal stripping during an orbit around a pure NFW host is insufficient to reconcile a `too big to fail' satellite with observed dSphs. Our NR dwarf does not agree with the half-light velocities of the LG dSph population after 5~Gyr of evolution in circular or elliptical orbits.

The combination of tidal evolution around a MW with both an NFW and disc component and baryonic removal with adiabatic expansion was tested with the exponential and instantaneous mass removal dwarfs. Both experienced large amounts of mass removal and a significant reduction in $\vmax$. This brought them into agreement with every LG dSph over the course of at least one of the elliptical or circular orbits.

Tidal evolution alone was also examined in our pre-halo formation mass removal dwarf. This subhalo had the same mass as our other baryon removal cases but without adiabatic expansion. It experienced weaker tidal stripping due to its large inner density profile, making it compatible only with the five most massive LG dSph population after evolution in elliptical and circular orbits.

In order to quantify how the inner regions of the dwarf are impacted due to tidal stripping, we created two analytical models. The first calculated how a dwarf's tidal radius changes depending on the mass of the dwarf, the mass of the primary, the distance from the main halo and the inclusion or absence of baryons in the central region of the host. We found that $\rtrs$, the tidal radius over the scale radius of the satellite, is able to predict the strength of tidal stripping. When $\rtrs$ is greater than 4, there is no significant evolution in $\vmax$. However, when $\rtrs$ is less than 2, the central region of the dwarf is subject to tidal forces and may be significantly altered.

The second analytical model for tidal stripping iteratively removed unbound particles, estimated the correct adiabatic expansion the halo would undergo from this amount of removal and then recalculated the tidal radius. This model was able to reproduce our {\it N}-body simulations in the context of reduction in $\vmax$ and mass as a function of $\rtrs$ and time in orbit. We find that each iteration continues to remove some fraction of the particles without convergence on a single value.\\ 

Our main conclusions are as follows.
\begin{enumerate}
\item Our simulations demonstrate that pure {\it N}-body simulations, which incorrectly assume a universal fraction of baryons everywhere, are unable to match observations of LG dSphs, as found in the \citet{BoylanKolchin:2012id} analysis of the Aquarius simulations.
\item However, inclusion of baryon removal (of the order of 20 per cent) with adiabatic expansion is very effective, bringing initially extremely massive dwarfs into agreement with observations before any orbital evolution.
\item Including the effects of tidal stripping without expansion due to baryon removal is insufficient to reproduce the entire observed dSph population for the case of an initially massive dwarf.
\item The inclusion of both baryon removal with adiabatic expansion {\it and} tidal stripping is extremely powerful, capable of transforming an initially incompatible dwarf into one that agrees with the entire LG dSph population.
\item Dwarfs undergo continual mass removal and reduction in $\vmax$ at each pericentre passage. These values do not converge even after 6~Gyr of evolution.
\item Tidal radius over scale radius, $\rtrs$, is an important parameter for understanding the potential mass and $\vmax$ loss of a dwarf. If $\rtrs$ is less than 4, satellites may become substantially disrupted, depending on the number of times they orbit their host.
\end{enumerate}

\section*{Acknowledgements}
The authors would like to thank Andrey Kravtsov for his helpful comments on an earlier version of this paper.
Additional thanks goes to the anonymous referee for their comments, which improved the clarity of this document. K.A. was supported by the National Science Foundation under Grant No. DGE-1144468. S.T. and A.K. were supported by grant STSci/HST-AR-12647.01 and the collaborative grant NSF-AST-1009908. Some simulations were performed at the Leibniz Rechenzentrum Munich (LRZ) and others were performed at the National Energy Research Scientific Computing Center, which is supported by the Office of Science of the U.S. Department of Energy under Contract No. DE-AC02-05CH11231.

\footnotesize{
\bibliographystyle{mn2e}
\setlength{\bibhang}{2.0em}
\setlength\labelwidth{0.0em}
\bibliography{paper_refs}

\providecommand{\noopsort}[1]{}
\begin{thebibliography}{115}
\expandafter\ifx\csname natexlab\endcsname\relax\def\natexlab#1{#1}\fi

\bibitem[{Agertz {et~al}\mbox{.}(2013)Agertz, Kravtsov, Leitner, \&
  Gnedin}]{Agertz:2013il}
Agertz O., Kravtsov A.~V., Leitner S.~N., Gnedin N.~Y., 2013, ApJ, 770, 25

\bibitem[{Amorisco \& Evans(2012)}]{Amorisco:2012fp}
Amorisco N.~C., Evans N.~W., 2012, MNRAS, 419, 184

\bibitem[{Benson {et~al}\mbox{.}(2002)Benson, Frenk, Lacey, Baugh, \&
  Cole}]{Benson:2002ek}
Benson A.~J., Frenk C.~S., Lacey C.~G., Baugh C.~M., Cole S., 2002, MNRAS, 333,
  177

\bibitem[{Blumenthal {et~al}\mbox{.}(1986)Blumenthal, Faber, Flores, \&
  Primack}]{Blumenthal:1986ie}
Blumenthal G.~R., Faber S.~M., Flores R., Primack J.~R., 1986, ApJ, 301, 27

\bibitem[{Bovy {et~al}\mbox{.}(2012)Bovy, Allende~Prieto, Beers, Bizyaev,
  da~Costa, Cunha, Ebelke, Eisenstein, Frinchaboy, Garc{\'\i}a~P{\'e}rez,
  Girardi, Hearty, Hogg, Holtzman, Maia, Majewski, Malanushenko, Malanushenko,
  M{\'e}sz{\'a}ros, Nidever, O'Connell, O'Donnell, Oravetz, Pan, Rocha-Pinto,
  Schiavon, Schneider, Schultheis, Skrutskie, Smith, Weinberg, Wilson, \&
  Zasowski}]{Bovy:2012gj}
Bovy J. {et~al.}, 2012, ApJ, 759, 131

\bibitem[{Boylan-Kolchin {et~al}\mbox{.}(2011)Boylan-Kolchin, Bullock, \&
  Kaplinghat}]{BoylanKolchin:2011ky}
Boylan-Kolchin M., Bullock J.~S., Kaplinghat M., 2011, MNRAS, 415, L40

\bibitem[{Boylan-Kolchin {et~al}\mbox{.}(2012)Boylan-Kolchin, Bullock, \&
  Kaplinghat}]{BoylanKolchin:2012id}
Boylan-Kolchin M., Bullock J.~S., Kaplinghat M., 2012, MNRAS, 422, 1203

\bibitem[{Boylan-Kolchin {et~al}\mbox{.}(2013)Boylan-Kolchin, Bullock, Sohn,
  Besla, \& van~der Marel}]{BoylanKolchin:2013et}
Boylan-Kolchin M., Bullock J.~S., Sohn S.~T., Besla G., van~der Marel R.~P.,
  2013, ApJ, 768, 140

\bibitem[{Boylan-Kolchin {et~al}\mbox{.}(2010)Boylan-Kolchin, Springel, White,
  \& Jenkins}]{BoylanKolchin:2010bd}
Boylan-Kolchin M., Springel V., White S. D.~M., Jenkins A., 2010, MNRAS, 406,
  896

\bibitem[{Brooks \& Zolotov(2012)}]{Brooks:2012wa}
Brooks A.~M., Zolotov A., 2012, preprint (arXiv:1207.2468)

\bibitem[{Bullock {et~al}\mbox{.}(2000)Bullock, Kravtsov, \&
  Weinberg}]{Bullock:2000bn}
Bullock J.~S., Kravtsov A.~V., Weinberg D.~H., 2000, ApJ, 539, 517

\bibitem[{Bullock {et~al}\mbox{.}(2010)Bullock, Stewart, Kaplinghat, Tollerud,
  \& Wolf}]{Bullock:2010jz}
Bullock J.~S., Stewart K.~R., Kaplinghat M., Tollerud E.~J., Wolf J., 2010,
  ApJ, 717, 1043

\bibitem[{Busha {et~al}\mbox{.}(2011{\natexlab{a}})Busha, Marshall, Wechsler,
  Klypin, \& Primack}]{Busha:2011jp}
Busha M.~T., Marshall P.~J., Wechsler R.~H., Klypin A., Primack J.,
  2011{\natexlab{a}}, ApJ, 743, 40

\bibitem[{Busha {et~al}\mbox{.}(2011{\natexlab{b}})Busha, Wechsler, Behroozi,
  Gerke, Klypin, \& Primack}]{Busha:2011gk}
Busha M.~T., Wechsler R.~H., Behroozi P.~S., Gerke B.~F., Klypin A.~A., Primack
  J.~R., 2011{\natexlab{b}}, ApJ, 743, 117

\bibitem[{Ceverino \& Klypin(2009)}]{Ceverino:2009ke}
Ceverino D., Klypin A., 2009, ApJ, 695, 292

\bibitem[{Col{\'\i}n {et~al}\mbox{.}(2010)Col{\'\i}n, Avila-Reese,
  V{\'a}zquez-Semadeni, Valenzuela, \& Ceverino}]{Colin:2010ee}
Col{\'\i}n P., Avila-Reese V., V{\'a}zquez-Semadeni E., Valenzuela O., Ceverino
  D., 2010, ApJ, 713, 535

\bibitem[{Col{\'\i}n {et~al}\mbox{.}(2006)Col{\'\i}n, Valenzuela, \&
  Klypin}]{Colin:2006fp}
Col{\'\i}n P., Valenzuela O., Klypin A., 2006, ApJ, 644, 687

\bibitem[{Cuesta {et~al}\mbox{.}(2008)Cuesta, Prada, Klypin, \&
  Moles}]{Cuesta:2008gl}
Cuesta A.~J., Prada F., Klypin A., Moles M., 2008, MNRAS, 389, 385

\bibitem[{de~Blok \& Bosma(2002)}]{deBlok:2002bw}
de~Blok W. J.~G., Bosma A., 2002, A{\&}A, 385, 816

\bibitem[{de~Blok {et~al}\mbox{.}(2001)de~Blok, McGaugh, Bosma, \&
  Rubin}]{deBlok:2001hs}
de~Blok W. J.~G., McGaugh S.~S., Bosma A., Rubin V.~C., 2001, ApJ, 552, L23

\bibitem[{Dehnen \& Binney(1998)}]{Dehnen:1998gx}
Dehnen W., Binney J., 1998, MNRAS, 294, 429

\bibitem[{Dekel {et~al}\mbox{.}(2003{\natexlab{a}})Dekel, {\noopsort{a}}Devor,
  \& Hetzroni}]{Dekel:2003gu}
Dekel A., {\noopsort{a}}Devor J., Hetzroni G., 2003{\natexlab{a}}, MNRAS, 341,
  326

\bibitem[{Dekel {et~al}\mbox{.}(2003{\natexlab{b}})Dekel, {\noopsort{b}}Arad,
  Devor, \& Birnboim}]{Dekel:2003ew}
Dekel A., {\noopsort{b}}Arad I., Devor J., Birnboim Y., 2003{\natexlab{b}},
  ApJ, 588, 680

\bibitem[{Dekel \& Silk(1986)}]{Dekel:1986cv}
Dekel A., Silk J., 1986, ApJ, 303, 39

\bibitem[{Dekel \& Woo(2003)}]{Dekel:2003id}
Dekel A., Woo J., 2003, MNRAS, 344, 1131

\bibitem[{Di~Cintio {et~al}\mbox{.}(2013)Di~Cintio, Knebe, Libeskind, Brook,
  Yepes, Gottl{\"o}ber, \& Hoffman}]{DiCintio:2013fo}
Di~Cintio A., Knebe A., Libeskind N.~I., Brook C., Yepes G., Gottl{\"o}ber S.,
  Hoffman Y., 2013, MNRAS, 431, 1220

\bibitem[{Di~Cintio {et~al}\mbox{.}(2011)Di~Cintio, Knebe, Libeskind, Yepes,
  Gottl{\"o}ber, \& Hoffman}]{DiCintio:2011df}
Di~Cintio A., Knebe A., Libeskind N.~I., Yepes G., Gottl{\"o}ber S., Hoffman
  Y., 2011, MNRAS, 417, L74

\bibitem[{Diemand {et~al}\mbox{.}(2007)Diemand, Kuhlen, \&
  Madau}]{Diemand:2007hb}
Diemand J., Kuhlen M., Madau P., 2007, ApJ, 667, 859

\bibitem[{Diemand {et~al}\mbox{.}(2008)Diemand, Kuhlen, Madau, Zemp, Moore,
  Potter, \& Stadel}]{Diemand:2008hr}
Diemand J., Kuhlen M., Madau P., Zemp M., Moore B., Potter D., Stadel J., 2008,
  Nature, 454, 735

\bibitem[{Diemer {et~al}\mbox{.}(2013)Diemer, More, \&
  Kravtsov}]{Diemer:2013hu}
Diemer B., More S., Kravtsov A.~V., 2013, ApJ, 766, 25

\bibitem[{Einasto {et~al}\mbox{.}(1974)Einasto, Saar, Kaasik, \&
  Chernin}]{Einasto:1974fh}
Einasto J., Saar E., Kaasik A., Chernin A.~D., 1974, Nature, 252, 111

\bibitem[{Ferrero {et~al}\mbox{.}(2012)Ferrero, Abadi, Navarro, Sales, \&
  Gurovich}]{Ferrero:2012bt}
Ferrero I., Abadi M.~G., Navarro J.~F., Sales L.~V., Gurovich S., 2012, MNRAS,
  425, 2817

\bibitem[{Flynn {et~al}\mbox{.}(2006)Flynn, Holmberg, Portinari, Fuchs, \&
  Jahrei{\ss}}]{Flynn:2006he}
Flynn C., Holmberg J., Portinari L., Fuchs B., Jahrei{\ss} H., 2006, MNRAS,
  372, 1149

\bibitem[{Gerhard(2002)}]{Gerhard:2002wx}
Gerhard O., 2002, Space Sci. Rev., 100, 129

\bibitem[{Ghigna {et~al}\mbox{.}(1998)Ghigna, Moore, Governato, Lake, Quinn, \&
  Stadel}]{Ghigna:1998ik}
Ghigna S., Moore B., Governato F., Lake G., Quinn T., Stadel J., 1998, MNRAS,
  300, 146

\bibitem[{Gnedin {et~al}\mbox{.}(2011)Gnedin, Ceverino, Gnedin, Klypin,
  Kravtsov, Levine, Nagai, \& Yepes}]{Gnedin:2011tf}
Gnedin O.~Y., Ceverino D., Gnedin N.~Y., Klypin A.~A., Kravtsov A.~V., Levine
  R., Nagai D., Yepes G., 2011, preprint (arXiv:1108.5736)

\bibitem[{Gnedin {et~al}\mbox{.}(2004)Gnedin, Kravtsov, Klypin, \&
  Nagai}]{Gnedin:2004hc}
Gnedin O.~Y., Kravtsov A.~V., Klypin A.~A., Nagai D., 2004, ApJ, 616, 16

\bibitem[{Gnedin \& Zhao(2002)}]{Gnedin:2002bu}
Gnedin O.~Y., Zhao H., 2002, MNRAS, 333, 299

\bibitem[{Governato {et~al}\mbox{.}(2010)Governato, Brook, Mayer, Brooks, Rhee,
  Wadsley, Jonsson, Willman, Stinson, Quinn, \& Madau}]{Governato:2010ed}
Governato F. {et~al.}, 2010, Nature, 463, 203

\bibitem[{Governato {et~al}\mbox{.}(2007)Governato, Willman, Mayer, Brooks,
  Stinson, Valenzuela, Wadsley, \& Quinn}]{Governato:2007dq}
Governato F., Willman B., Mayer L., Brooks A., Stinson G., Valenzuela O.,
  Wadsley J., Quinn T., 2007, MNRAS, 374, 1479

\bibitem[{Governato {et~al}\mbox{.}(2012)Governato, Zolotov, Pontzen,
  Christensen, Oh, Brooks, Quinn, Shen, \& Wadsley}]{Governato:2012cw}
Governato F. {et~al.}, 2012, MNRAS, 422, 1231

\bibitem[{Grebel(1998)}]{Grebel:1998uz}
Grebel E.~K., 1998, Moriond Astrophysics Meetings, Dwarf Galaxies and Cosmology
  Editions Fronti{\`e}res, Paris, p. 125

\bibitem[{Guo {et~al}\mbox{.}(2010)Guo, White, Li, \&
  Boylan-Kolchin}]{Guo:2010do}
Guo Q., White S., Li C., Boylan-Kolchin M., 2010, MNRAS, 404, 1111

\bibitem[{Gustafsson {et~al}\mbox{.}(2006)Gustafsson, Fairbairn, \&
  Sommer-Larsen}]{Gustafsson:2006fr}
Gustafsson M., Fairbairn M., Sommer-Larsen J., 2006, Phys. Rev. D, 74, 123522

\bibitem[{Hayashi {et~al}\mbox{.}(2003)Hayashi, Navarro, Taylor, Stadel, \&
  Quinn}]{Hayashi:2003kz}
Hayashi E., Navarro J.~F., Taylor J.~E., Stadel J., Quinn T., 2003, ApJ, 584,
  541

\bibitem[{Hayashi \& Chiba(2012)}]{Hayashi:2012fn}
Hayashi K., Chiba M., 2012, ApJ, 755, 145

\bibitem[{Hopkins {et~al}\mbox{.}(2011)Hopkins, Quataert, \&
  Murray}]{Hopkins:2011fk}
Hopkins P.~F., Quataert E., Murray N., 2011, MNRAS, 417, 950

\bibitem[{Kallivayalil {et~al}\mbox{.}(2009)Kallivayalil, Besla, Sanderson, \&
  Alcock}]{Kallivayalil:2009ho}
Kallivayalil N., Besla G., Sanderson R., Alcock C., 2009, ApJ, 700, 924

\bibitem[{Karachentsev(2005)}]{Karachentsev:2005gb}
Karachentsev I.~D., 2005, AJ, 129, 178

\bibitem[{Karachentsev {et~al}\mbox{.}(2005)Karachentsev, Karachentseva, \&
  Sharina}]{Karachentsev:2005jn}
Karachentsev I.~D., Karachentseva V.~E., Sharina M.~E., 2005, Proc. IAU Symp.
  198, Near-Fields Cosmology with Dwarf Elliptical Galaxies. Cambridge Univ.
  Press, Cambridge, p. 295

\bibitem[{Klypin {et~al}\mbox{.}(1999{\natexlab{a}})Klypin, Gottl{\"o}ber,
  Kravtsov, \& Khokhlov}]{Klypin:1999bk}
Klypin A., Gottl{\"o}ber S., Kravtsov A.~V., Khokhlov A.~M.,
  1999{\natexlab{a}}, ApJ, 516, 530

\bibitem[{Klypin {et~al}\mbox{.}(1999{\natexlab{b}})Klypin, Kravtsov,
  Valenzuela, \& Prada}]{Klypin:1999ej}
Klypin A., Kravtsov A.~V., Valenzuela O., Prada F., 1999{\natexlab{b}}, ApJ,
  522, 82

\bibitem[{Klypin {et~al}\mbox{.}(2002)Klypin, Zhao, \&
  Somerville}]{Klypin:2002bm}
Klypin A., Zhao H., Somerville R.~S., 2002, ApJ, 573, 597

\bibitem[{Komatsu {et~al}\mbox{.}(2011)Komatsu, Smith, Dunkley, Bennett, Gold,
  Hinshaw, Jarosik, Larson, Nolta, Page, Spergel, Halpern, Hill, Kogut, Limon,
  Meyer, Odegard, Tucker, Weiland, Wollack, \& Wright}]{Komatsu:2011in}
Komatsu E. {et~al.}, 2011, ApJS, 192, 18

\bibitem[{Koposov {et~al}\mbox{.}(2008)Koposov, Belokurov, Evans, {Hewett, P.
  C.}, Irwin, Gilmore, Zucker, Rix, Fellhauer, Bell, \&
  Glushkova}]{Koposov:2008ja}
Koposov S. {et~al.}, 2008, ApJ, 686, 279

\bibitem[{Koposov {et~al}\mbox{.}(2009)Koposov, Yoo, Rix, Weinberg, Macci{\`o},
  \& Escud{\'e}}]{Koposov:2009kh}
Koposov S.~E., Yoo J., Rix H.-W., Weinberg D.~H., Macci{\`o} A.~V., Escud{\'e}
  J.~M., 2009, ApJ, 696, 2179

\bibitem[{Kormendy(1985)}]{Kormendy:1985bl}
Kormendy J., 1985, ApJ, 295, 73

\bibitem[{Kravtsov(1999)}]{Kravtsov:1999vi}
Kravtsov A.~V., 1999, PhD thesis, New Mexico State University

\bibitem[{Kravtsov {et~al}\mbox{.}(2004)Kravtsov, Gnedin, \&
  Klypin}]{Kravtsov:2004he}
Kravtsov A.~V., Gnedin O.~Y., Klypin A.~A., 2004, ApJ, 609, 482

\bibitem[{Kravtsov {et~al}\mbox{.}(1997)Kravtsov, Klypin, \&
  Khokhlov}]{Kravtsov:1997iy}
Kravtsov A.~V., Klypin A.~A., Khokhlov A.~M., 1997, ApJS, 111, 73

\bibitem[{Leitherer {et~al}\mbox{.}(1999)Leitherer, Schaerer, Goldader,
  Gonz{\'a}lez~Delgado, Robert, Kune, de~Mello, Devost, \&
  Heckman}]{Leitherer:1999jt}
Leitherer C. {et~al.}, 1999, ApJS, 123, 3

\bibitem[{L{\'e}pine {et~al}\mbox{.}(2011)L{\'e}pine, Koch, Rich, \&
  Kuijken}]{Lepine:2011if}
L{\'e}pine S., Koch A., Rich R.~M., Kuijken K., 2011, ApJ, 741, 100

\bibitem[{Li \& White(2008)}]{Li:2008ji}
Li Y.-S., White S. D.~M., 2008, MNRAS, 384, 1459

\bibitem[{Lin \& Faber(1983)}]{Lin:1983ha}
Lin D. N.~C., Faber S.~M., 1983, ApJ, 266, L21

\bibitem[{Lovell {et~al}\mbox{.}(2012)Lovell, Eke, Frenk, Gao, Jenkins, Theuns,
  Wang, White, Boyarsky, \& Ruchayskiy}]{Lovell:2012gp}
Lovell M.~R. {et~al.}, 2012, MNRAS, 420, 2318

\bibitem[{Mashchenko {et~al}\mbox{.}(2006)Mashchenko, Couchman, \&
  Wadsley}]{Mashchenko:2006ff}
Mashchenko S., Couchman H. M.~P., Wadsley J., 2006, Nature, 442, 539

\bibitem[{Mashchenko {et~al}\mbox{.}(2008)Mashchenko, Wadsley, \&
  Couchman}]{Mashchenko:2008fa}
Mashchenko S., Wadsley J., Couchman H. M.~P., 2008, Sci, 319, 174

\bibitem[{McConnachie(2012)}]{McConnachie:2012fh}
McConnachie A.~W., 2012, AJ, 144, 4

\bibitem[{Moore {et~al}\mbox{.}(1999)Moore, Ghigna, Governato, Lake, Quinn,
  Stadel, \& Tozzi}]{Moore:1999ja}
Moore B., Ghigna S., Governato F., Lake G., Quinn T., Stadel J., Tozzi P.,
  1999, ApJ, 524, L19

\bibitem[{Navarro {et~al}\mbox{.}(1996{\natexlab{a}})Navarro, Eke, \&
  Frenk}]{Navarro:1996ws}
Navarro J.~F., Eke V.~R., Frenk C.~S., 1996{\natexlab{a}}, MNRAS, 283, L72

\bibitem[{Navarro {et~al}\mbox{.}(1996{\natexlab{b}})Navarro, Frenk, \&
  White}]{Navarro:1996ce}
Navarro J.~F., Frenk C.~S., White S. D.~M., 1996{\natexlab{b}}, ApJ, 462, 563

\bibitem[{Navarro {et~al}\mbox{.}(1997)Navarro, Frenk, \&
  White}]{Navarro:1997if}
Navarro J.~F., Frenk C.~S., White S. D.~M., 1997, ApJ, 490, 493

\bibitem[{Navarro {et~al}\mbox{.}(2010)Navarro, Ludlow, Springel, Wang,
  Vogelsberger, White, Jenkins, Frenk, \& Helmi}]{Navarro:2010hn}
Navarro J.~F. {et~al.}, 2010, MNRAS, 402, 21

\bibitem[{Oh {et~al}\mbox{.}(2011)Oh, de~Blok, Brinks, Walter, \&
  Kennicutt}]{Oh:2011fk}
Oh S.-H., de~Blok W. J.~G., Brinks E., Walter F., Kennicutt R. C.~J., 2011, AJ,
  141, 193

\bibitem[{Okamoto \& Frenk(2009)}]{Okamoto:2009jd}
Okamoto T., Frenk C.~S., 2009, MNRAS, 399, L174

\bibitem[{Okamoto {et~al}\mbox{.}(2008)Okamoto, Gao, \&
  Theuns}]{Okamoto:2008ha}
Okamoto T., Gao L., Theuns T., 2008, MNRAS, 390, 920

\bibitem[{Parry {et~al}\mbox{.}(2012)Parry, Eke, Frenk, \&
  Okamoto}]{Parry:2012fd}
Parry O.~H., Eke V.~R., Frenk C.~S., Okamoto T., 2012, MNRAS, 419, 3304

\bibitem[{Pe{\~n}arrubia {et~al}\mbox{.}(2010)Pe{\~n}arrubia, Benson, Walker,
  Gilmore, McConnachie, \& Mayer}]{Penarrubia:2010du}
Pe{\~n}arrubia J., Benson A.~J., Walker M.~G., Gilmore G., McConnachie A.~W.,
  Mayer L., 2010, MNRAS, 406, 1290

\bibitem[{Pe{\~n}arrubia {et~al}\mbox{.}(2008)Pe{\~n}arrubia, Navarro, \&
  McConnachie}]{Penarrubia:2008eq}
Pe{\~n}arrubia J., Navarro J.~F., McConnachie A.~W., 2008, ApJ, 673, 226

\bibitem[{Pe{\~n}arrubia {et~al}\mbox{.}(2009)Pe{\~n}arrubia, Navarro,
  McConnachie, \& Martin}]{Penarrubia:2009kp}
Pe{\~n}arrubia J., Navarro J.~F., McConnachie A.~W., Martin N.~F., 2009, ApJ,
  698, 222

\bibitem[{Pe{\~n}arrubia {et~al}\mbox{.}(2012)Pe{\~n}arrubia, Pontzen, Walker,
  \& Koposov}]{Penarrubia:2012dy}
Pe{\~n}arrubia J., Pontzen A., Walker M.~G., Koposov S.~E., 2012, ApJ, 759, L42

\bibitem[{Peter {et~al}\mbox{.}(2013)Peter, Rocha, Bullock, \&
  Kaplinghat}]{Peter:2013iu}
Peter A. H.~G., Rocha M., Bullock J.~S., Kaplinghat M., 2013, MNRAS, 430, 105

\bibitem[{Piatek {et~al}\mbox{.}(2005)Piatek, Pryor, Bristow, Olszewski,
  Harris, Mateo, Minniti, \& Tinney}]{Piatek:2005ii}
Piatek S., Pryor C., Bristow P., Olszewski E.~W., Harris H.~C., Mateo M.,
  Minniti D., Tinney C.~G., 2005, AJ, 130, 95

\bibitem[{Piatek {et~al}\mbox{.}(2006)Piatek, Pryor, Bristow, Olszewski,
  Harris, Mateo, Minniti, \& Tinney}]{Piatek:2006jz}
Piatek S., Pryor C., Bristow P., Olszewski E.~W., Harris H.~C., Mateo M.,
  Minniti D., Tinney C.~G., 2006, AJ, 131, 1445

\bibitem[{Piatek {et~al}\mbox{.}(2007)Piatek, Pryor, Bristow, Olszewski,
  Harris, Mateo, Minniti, \& Tinney}]{Piatek:2007ht}
Piatek S., Pryor C., Bristow P., Olszewski E.~W., Harris H.~C., Mateo M.,
  Minniti D., Tinney C.~G., 2007, AJ, 133, 818

\bibitem[{Piatek {et~al}\mbox{.}(2003)Piatek, Pryor, Olszewski, Harris, Mateo,
  Minniti, \& Tinney}]{Piatek:2003dd}
Piatek S., Pryor C., Olszewski E.~W., Harris H.~C., Mateo M., Minniti D.,
  Tinney C.~G., 2003, AJ, 126, 2346

\bibitem[{Pontzen \& Governato(2012)}]{Pontzen:2012jg}
Pontzen A., Governato F., 2012, MNRAS, 421, 3464

\bibitem[{Quinn {et~al}\mbox{.}(1996)Quinn, Katz, \& Efstathiou}]{Quinn:1996wd}
Quinn T., Katz N., Efstathiou G., 1996, MNRAS, 278, L49

\bibitem[{Reed {et~al}\mbox{.}(2005)Reed, Governato, Verde, Gardner, Quinn,
  Stadel, Merritt, \& Lake}]{Reed:2005fe}
Reed D., Governato F., Verde L., Gardner J., Quinn T., Stadel J., Merritt D.,
  Lake G., 2005, MNRAS, 357, 82

\bibitem[{Rocha {et~al}\mbox{.}(2013)Rocha, Peter, Bullock, Kaplinghat,
  Garrison-Kimmel, O{\~n}orbe, \& Moustakas}]{Rocha:2013bo}
Rocha M., Peter A. H.~G., Bullock J.~S., Kaplinghat M., Garrison-Kimmel S.,
  O{\~n}orbe J., Moustakas L.~A., 2013, MNRAS, 430, 81

\bibitem[{Sakamoto {et~al}\mbox{.}(2003)Sakamoto, Chiba, \&
  Beers}]{Sakamoto:2003jw}
Sakamoto T., Chiba M., Beers T.~C., 2003, A{\&}A, 397, 899

\bibitem[{Smith {et~al}\mbox{.}(2007)Smith, Ruchti, Helmi, Wyse, Fulbright,
  Freeman, Navarro, Seabroke, Steinmetz, Williams, Bienayme, Binney,
  Bland-Hawthorn, Dehnen, Gibson, Gilmore, Grebel, Munari, Parker, Scholz,
  Siebert, Watson, \& Zwitter}]{Smith:2007fi}
Smith M.~C. {et~al.}, 2007, MNRAS, 379, 755

\bibitem[{Sohn {et~al}\mbox{.}(2013)Sohn, Besla, van~der Marel, Boylan-Kolchin,
  Majewski, \& Bullock}]{Sohn:2013do}
Sohn S.~T., Besla G., van~der Marel R.~P., Boylan-Kolchin M., Majewski S.~R.,
  Bullock J.~S., 2013, ApJ, 768, 139

\bibitem[{Somerville(2002)}]{Somerville:2002ky}
Somerville R.~S., 2002, ApJ, 572, L23

\bibitem[{Springel {et~al}\mbox{.}(2008)Springel, Wang, Vogelsberger, Ludlow,
  Jenkins, Helmi, Navarro, Frenk, \& White}]{Springel:2008gd}
Springel V. {et~al.}, 2008, MNRAS, 391, 1685

\bibitem[{Strigari {et~al}\mbox{.}(2006)Strigari, Bullock, Kaplinghat,
  Kravtsov, Gnedin, Abazajian, \& Klypin}]{Strigari:2006ik}
Strigari L.~E., Bullock J.~S., Kaplinghat M., Kravtsov A.~V., Gnedin O.~Y.,
  Abazajian K., Klypin A.~A., 2006, ApJ, 652, 306

\bibitem[{Strigari {et~al}\mbox{.}(2010)Strigari, Frenk, \&
  White}]{Strigari:2010kl}
Strigari L.~E., Frenk C.~S., White S. D.~M., 2010, MNRAS, 408, 2364

\bibitem[{Swaters {et~al}\mbox{.}(2003)Swaters, Madore, van~den Bosch, \&
  Balcells}]{Swaters:2003kf}
Swaters R.~A., Madore B.~F., van~den Bosch F.~C., Balcells M., 2003, ApJ, 583,
  732

\bibitem[{Tikhonov \& Klypin(2009)}]{Tikhonov:2009eh}
Tikhonov A.~V., Klypin A., 2009, MNRAS, 395, 1915

\bibitem[{Tissera {et~al}\mbox{.}(2010)Tissera, White, Pedrosa, \&
  Scannapieco}]{Tissera:2010ka}
Tissera P.~B., White S. D.~M., Pedrosa S., Scannapieco C., 2010, MNRAS, 406,
  922

\bibitem[{Tollerud {et~al}\mbox{.}(2008)Tollerud, Bullock, Strigari, \&
  Willman}]{Tollerud:2008eq}
Tollerud E.~J., Bullock J.~S., Strigari L.~E., Willman B., 2008, ApJ, 688, 277

\bibitem[{Trujillo-Gomez {et~al}\mbox{.}(2013)Trujillo-Gomez, Klypin,
  Col{\'\i}n, Ceverino, Arraki, \& Primack}]{TrujilloGomez:2013ts}
Trujillo-Gomez S., Klypin A., Col{\'\i}n P., Ceverino D., Arraki K., Primack
  J., 2013, preprint (arXiv:1311.2910), 2910

\bibitem[{Trujillo-Gomez {et~al}\mbox{.}(2011)Trujillo-Gomez, Klypin, Primack,
  \& Romanowsky}]{TrujilloGomez:2011js}
Trujillo-Gomez S., Klypin A., Primack J., Romanowsky A.~J., 2011, ApJ, 742, 16

\bibitem[{Vera-Ciro {et~al}\mbox{.}(2013)Vera-Ciro, Helmi, Starkenburg, \&
  Breddels}]{VeraCiro:2013fy}
Vera-Ciro C.~A., Helmi A., Starkenburg E., Breddels M.~A., 2013, MNRAS, 428,
  1696

\bibitem[{Vogelsberger {et~al}\mbox{.}(2012)Vogelsberger, Zavala, \&
  Loeb}]{Vogelsberger:2012dy}
Vogelsberger M., Zavala J., Loeb A., 2012, MNRAS, 423, 3740

\bibitem[{Walker {et~al}\mbox{.}(2008)Walker, Mateo, \&
  Olszewski}]{Walker:2008cm}
Walker M.~G., Mateo M., Olszewski E.~W., 2008, ApJ, 688, L75

\bibitem[{Walker {et~al}\mbox{.}(2009)Walker, Mateo, Olszewski, Pe{\~n}arrubia,
  Wyn~Evans, \& Gilmore}]{Walker:2009ez}
Walker M.~G., Mateo M., Olszewski E.~W., Pe{\~n}arrubia J., Wyn~Evans N.,
  Gilmore G., 2009, ApJ, 704, 1274

\bibitem[{Walker \& Pe{\~n}arrubia(2011)}]{Walker:2011eg}
Walker M.~G., Pe{\~n}arrubia J., 2011, ApJ, 742, 20

\bibitem[{Wang {et~al}\mbox{.}(2012)Wang, Frenk, Navarro, Gao, \&
  Sawala}]{Wang:2012jg}
Wang J., Frenk C.~S., Navarro J.~F., Gao L., Sawala T., 2012, MNRAS, 424, 2715

\bibitem[{Watkins {et~al}\mbox{.}(2010)Watkins, Evans, \& An}]{Watkins:2010fc}
Watkins L.~L., Evans N.~W., An J.~H., 2010, MNRAS, 406, 264

\bibitem[{Wilkinson \& Evans(1999)}]{Wilkinson:1999cc}
Wilkinson M.~I., Evans N.~W., 1999, MNRAS, 310, 645

\bibitem[{Wolf \& Bullock(2012)}]{Wolf:2012vc}
Wolf J., Bullock J.~S., 2012, preprint (arXiv:1203.4240)

\bibitem[{Xue {et~al}\mbox{.}(2008)Xue, Rix, Zhao, Re~Fiorentin, Naab,
  Steinmetz, van~den Bosch, Beers, Lee, Bell, Rockosi, Yanny, Newberg, Wilhelm,
  Kang, Smith, \& Schneider}]{Xue:2008kb}
Xue X.~X. {et~al.}, 2008, ApJ, 684, 1143

\bibitem[{Zeldovich {et~al}\mbox{.}(1980)Zeldovich, Klypin, Khlopov, \&
  Checketkin}]{Zeldovich:1980ta}
Zeldovich Y.~B., Klypin A.~A., Khlopov M.~Y., Checketkin V.~M., 1980, Sov. J.
  Nucl. Phys., 31, 664

\bibitem[{Zolotov {et~al}\mbox{.}(2012)Zolotov, Brooks, Willman, Governato,
  Pontzen, Christensen, Dekel, Quinn, Shen, \& Wadsley}]{Zolotov:2012hi}
Zolotov A. {et~al.}, 2012, ApJ, 761, 71

\end{thebibliography}
}

\label{lastpage}
\end{document}